\documentclass[apj]{emulateapj}
\usepackage{graphicx}
\usepackage{epstopdf}
\usepackage{longtable}
\usepackage{graphicx}
\usepackage{amsmath}
\usepackage{amssymb}	
\usepackage{aas_macros}
\usepackage{journals}
\usepackage{footnote}

\newcommand{\HII}{\mbox{H II}}

\newcommand{\rion}[2]{{\ensuremath{\mbox{\rm #1$\,${\small\uppercase\expandafter{\romannumeral#2\relax}}}}}}

\shorttitle{Supernova Remnant Abundances}
\shortauthors{Dopita et al.}

\begin{document}

\title{Calibrating Interstellar Abundances using SNR Radiative Shocks}

\author{Michael A. Dopita \altaffilmark{1, 2}, Ivo R. Seitenzahl \altaffilmark{3,4}, Ralph S. Sutherland \altaffilmark{1, 2}, David C. Nicholls \altaffilmark{1, 2}, \\ Fr\'ed\'eric P.A. Vogt  \altaffilmark{5,6}, Parviz Ghavamian \altaffilmark{7}  \& Ashley J. Ruiter \altaffilmark{3,4}}

\altaffiltext{1}{Research School of Astronomy and Astrophysics, Australian National University, Cotter Road, Weston Creek, ACT 2611, Australia.}
\altaffiltext{2}{ARC Centre of Excellence for All Sky Astrophysics in 3 Dimensions (ASTRO 3D).}
\altaffiltext{3}{School of Physical, Environmental and Mathematical Sciences, University of New South Wales, Australian Defence Force Academy, Canberra, ACT 2600, Australia.}
\altaffiltext{4}{ARC Future Fellow.}
\altaffiltext{5}{European Southern Observatory, Av. Alonso de Cordova 3107, 763 0355 Vitacura, Santiago, Chile.}
\altaffiltext{6}{ESO Fellow.}
\altaffiltext{7}{Department of Physics, Astronomy and Geosciences, Towson University, Towson, MD, 21252.}

\begin{abstract}
Using integral field data we extract the optical spectra of shocked interstellar clouds in Kepler's supernova remnant located in the inner regions of our Galaxy,  as well as in the Large Magellanic Cloud (LMC), the Small Magellanic Cloud (SMC), NGC6822 and IC 1613. Using self-consistent shock modelling, we make a new determination of the chemical composition of the interstellar medium (ISM) in N, O, Ne, S, Cl and Ar in these galaxies and obtain accurate estimates of the fraction of refractory grains destroyed in the shock. By comparing our derived abundances with those obtained in recent works using observations of B stars, F supergiant stars and HII regions, we provide a new calibration for  abundance scaling in the range $7.9 \lesssim 12+\log {\mathrm {O/H}} \lesssim 9.1$.
\end{abstract}


\section{Introduction}\label{intro}
An accurate calibration of the extragalactic abundance scale is fundamental to our understanding of the chemical evolution of galaxies, since it provides a fossil record of the effect of generations of star formation, modulated by both inflows of inter-galactic gas, and outflows of heavy-element enhanced gas back into the inter-galactic gas pool. 

Typically, for high redshift studies, we must rely on ``metallicity'' calibrations based upon \HII\ regions. However, commonly used calibrations may disagree by more than an order of magnitude \citep{Kewley08}.  This discrepancy has many potential causes, including calibration errors resulting from reliance on measurements of electron temperature $T_e$  \citep{Pilyugin05,Pilyugin07},  hybrid approaches \citep{vanZee98}, relying only on strong emission line methods \citep{McGaugh91}, or modelling errors associated with  calibrating the theoretical photoionization models \citep{LopezSanchez12}. Other factors which may play a significant role in the abundance calibration using \HII\ regions are variations in gas pressure and ionisation parameter, the calibration of the N/O ratio at low metallicity and possible changes in the shape of the Initial Mass Function (IMF) \citep{Brinchmann08, Liu08, Kewley13a, Blanc15, Bian16, Bian18, Dopita16b}. 

Many of these issues, as well as the problem of the corrections due to the fraction of heavy elements locked in dust grains in the ISM, can be side-stepped by spectrophotometric observations of the photospheric abundances of B-stars \citep{Rolleston00, Smartt01, Korn02, Korn05, Hunter09, Takeda10, Nieva12}. These stars are sufficiently young to present ISM abundances at their surface, and are relatively simple to model compared with O-stars (which suffer from much stronger stellar winds, and modification of surface abundances from mixing processes such as meridional circulation). However, the intrinsic brightness of these stars limits detailed spectrophotometric studies to galaxies closer than the Virgo cluster.  The stellar calibration can be somewhat extended in distance by the use of F supergiants \citep{Russell92,Andrievsky01}, but there remains a question of whether the surface abundances of these stars has been modified by stellar evolution. Recently, \citet{Nicholls17} have combined both stellar and nebular abundances to provide an accurate estimate of the chemical abundances in the ISM close to the sun (the Local Galactic Concordance or LGC abundance set), and has provided simple scaling relations to fit stars in the halo of the Galaxy, and local galaxies.

A direct way to probe abundances in the ISM is to utilise radiative shocks propagating into dense, pristine interstellar clouds. Such shocks are to be found within supernova remnants (SNR). This technique was pioneered by \citet{Dopita76} and was further developed by \citet{Dopita80, Binette82} and \citet{Russell90}. These early models either did not include, or did not fully include, cooling and line radiation due to refractory elements such as Ca, Mg, Si, Fe and Ni. Furthermore, the return of these elements to the gas phase was not taken into account. 

Recently, \citet{Dopita16} investigated grain destruction across the Large Magellanic Cloud (LMC)  SNR N49 using integral field data, and determined that fast shocks ($v_s \gtrsim 250$\,km/s) appear to rapidly destroy their dust grains by classical  thermal sputtering, but slower shocks are ineffective in their grain destruction until the dense recombination zone is reached, at which point the dominant dust destruction process is the non-thermal sputtering and grain-grain collision mechanism advocated by \citet{Seab83} and \citet{Borkowski95}. These results were replicated in the more recent study of another LMC supernova remnant, N132D \citep{Dopita18}. 

Given that both the instrumental and analysis techniques have greatly improved since the pioneering \citet{Russell90} work, we were motivated to find to what extent the calibration of  radiative cloud shock spectroscopy in supernova remnants can complement the existing data from \HII\ regions, and from stellar spectroscopy of young stars, not only in our Galaxy, but in the Magellanic Clouds and Local Group dwarf galaxies. We present the results of this study here.

This paper is organised as follows. In Section \ref{sec:obs} we present the observational material and the extracted line fluxes. In Section \ref{sec:diagnostics} we present the methodology and results of our self-consistent shock analysis of these data. In Section \ref{abund} we present the derived abundances and the dust depletion factors for the ions of various refractory elements, while Section \ref{discuss} compares our new SNR calibration of the abundance scale with that derived previously using stellar or \HII\ region calibrations.

\section{Observations \& Data Reduction}\label{sec:obs}
The integral field spectra of these SNR were obtained using the Wide Field Spectrograph (WiFeS)  \citep{Dopita07,Dopita10}. This is an integral field spectrograph mounted on the  ANU 2.3m telescope at Siding Spring Observatory, which offers a field of view of 25\arcsec $\times$ 38\arcsec at a spatial resolution of either 1.0\arcsec $\times$ 0.5\arcsec or 1.0\arcsec $\times$ 1.0\arcsec, depending on the binning on the CCD. In these observations, we operated in the binned 1.0\arcsec x 1.0\arcsec\ mode. Most of the data were obtained in the low resolution mode $R \sim 3000$ (FWHM of $\sim 100$ km/s) using  the B3000 and R3000 gratings in each arm of the spectrograph, with the RT560 dichroic which provides a transition between the  two arms at around 560nm. This gives a continuous wavelength coverage from 3400--8900\AA. For observations of the two LMC SNR, N49 and N103B, the R7000 grating was used in the red (with the RT560 dichroic) giving a wavelength coverage only up to $\sim 7200$\AA. For details on the various observing modes available on WiFeS, see \citet{Dopita07}. 

The data for the LMC SNR, N49, N103B and N132D have already been described elsewhere \citep{Dopita16,Ghavamian17,Dopita18}. In Table \ref{Table1} we present the log of the observations on the remaining targets.

\begin{table*}
 \centering
   \caption{The log of  WiFeS observations of SNR}
    \label{Table1}
   \scalebox{0.95}{
  \begin{tabular}{lcccl}
 \hline
   Position & RA  & Dec & Date  & Exp. Time \\
& (J2000) &  (J2000) &   &   (s)  \\
   \hline \hline
IC 1613-S8 &  01:05:02.1 &  +02:08:43 &  20 Nov 2017 & $2\times1000$ \\
IC 1613-S8 &  01:05:02.1 &  +02:08:43 &  22 Nov 2017 & $2\times1200$ \\
SMC-0104-72.3 &  01:06:25.2 &  -72:05:29 &  28 Nov 2016 & $2\times1200$ \\
LMC-N103B & 05:08:58.6 & -68:43:33 & 18 Dec 2014 & $2\times 1800$ \\
LMC-N103B & 05:08:58.6 & -68:43:33 & 119 Dec 2014 & $2\times 1800$ \\
NGC6822-Ho12 & 19:44:56.5 & -14:48:30 & 7 Aug 2016 & $2\times1800$\\
NGC6822-Ho12 & 19:44:56.5 & -14:48:30 & 7 Aug 2016 & $1\times1800$\\
Kepler SNR & 17:30:36.2 & -21:28:49 & 13 May 2018 & $3\times500$ \\
Kepler SNR & 17:30:36.2 & -21:28:49 & 13 May 2018 & $2\times1500$ \\
 \hline
 \end{tabular}}
\end{table*}

The wavelength scale is calibrated using a series of Ne-Ar or Cu-Ar arc lamp exposures, taken throughout the night. Arc exposure times are 50s for the B3000 grating and 1s for the  R3000 grating. Flux calibration was performed using the STIS spectrophotometric standard stars  HD\,009051, HD\,031128, HD\,074000, HD\,111980 and HD\,200654\footnote{Available at : \newline {\url{www.mso.anu.edu.au/~bessell/FTP/Bohlin2013/GO12813.html}}}. For SNR Ho12 in NGC6822, HD\,128279 and CD-3018140 were used for flux calibration. In addition, B-type telluric standards HIP\,8352,  HIP\,41323 were observed to better correct for the OH and H$_2$O telluric absorption features in the red. The separation of these features by molecular species allows for a more accurate telluric correction by accounting for night to night variations in the column density of these two species. All data cubes were reduced using the PyWiFeS\footnote {\url{http://www.mso.anu.edu.au/pywifes/doku.php.}} data reduction pipeline \citep{Childress14,Childress14b}.

\subsection{Spectral Extraction}
\begin{figure}
 \centering
  \includegraphics[scale=0.4]{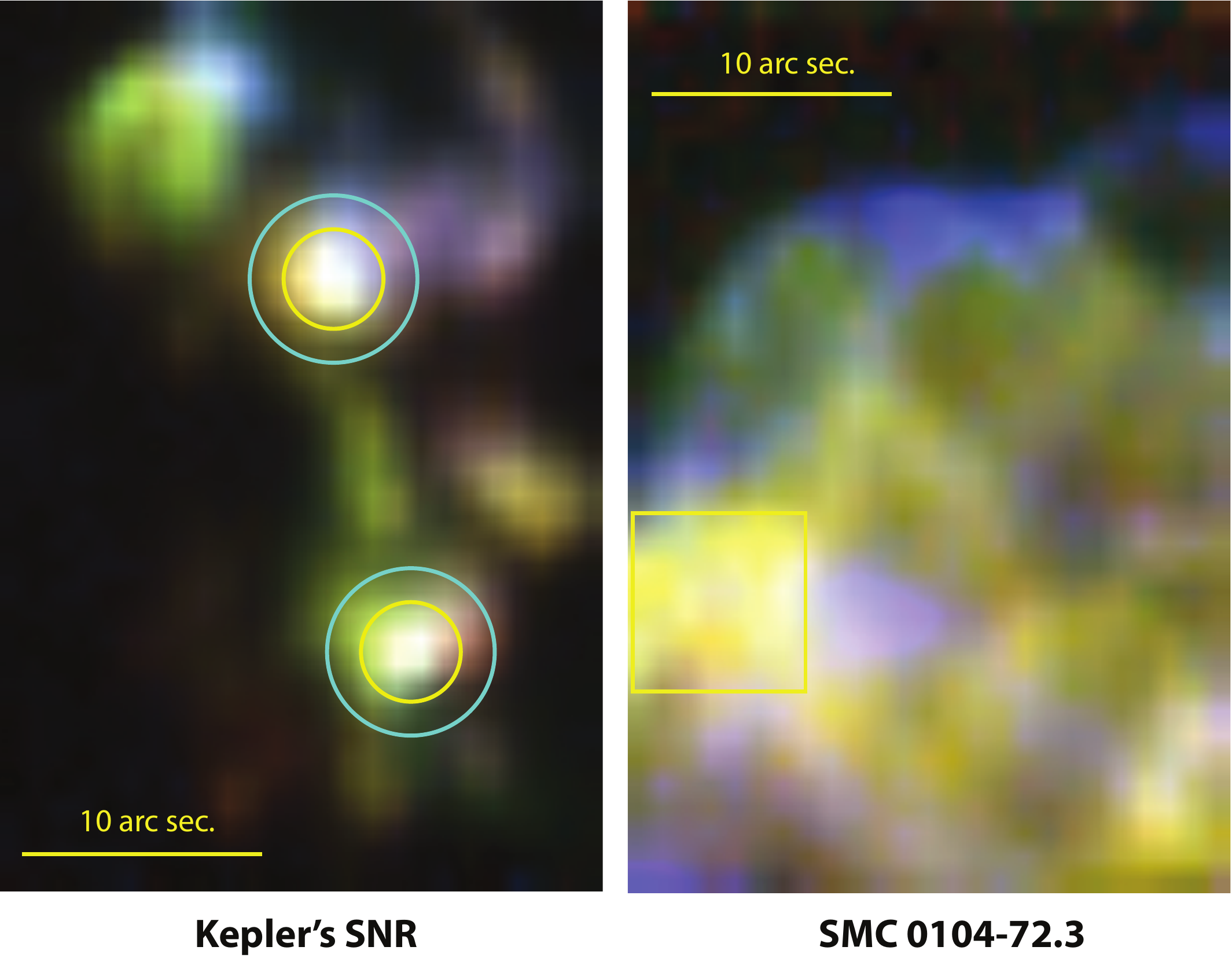}
  \caption{The extraction apertures for Kepler's  SNR and for SMC 0104-72.3, shown on the WiFeS images in [S II] (red), H$\alpha$ (green) and [O III] (blue). A square root stretch and boxcar smoothing has been applied to the image. For Kepler's SNR, North is at the top while for the SMC remnant North is to the right. For Kepler, the spectral extraction region is shown in yellow, while the sky subtraction annulus is located between the yellow and the turquoise circles. See \citet{Sankrit16} for HST images of this region. For the SMC remnant, the extraction region is shown as a yellow box, while the sky region comprised most of the dark spaxels lying outside the SNR at the top of the image.} \label{fig1}
 \end{figure}

\begin{figure}
 \centering
  \includegraphics[width =\columnwidth]{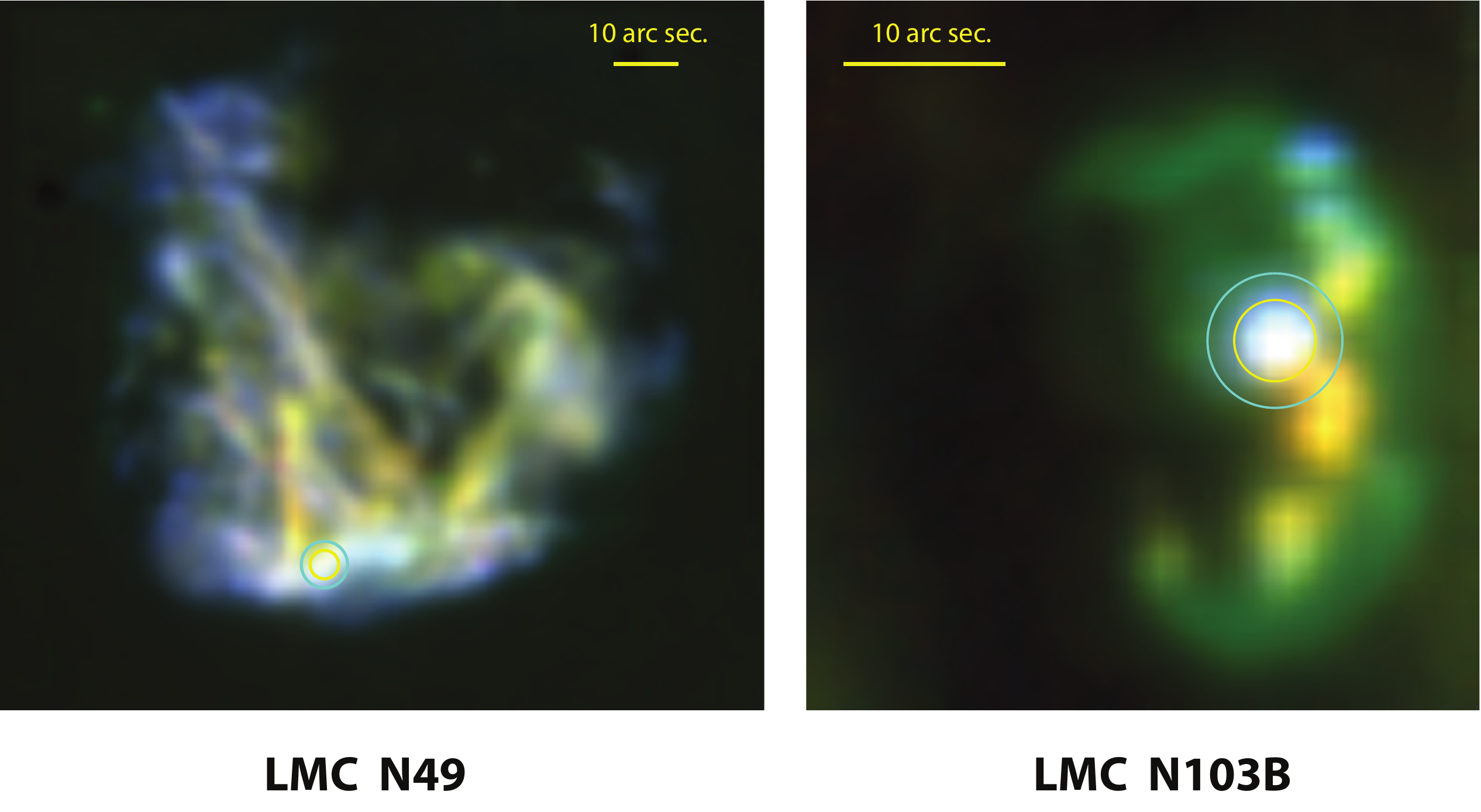}
  \caption{The extraction apertures for the LMC SNR, N49 (left) and N103B (right), shown on the WiFeS image mosaics formed from [S II] (red), H$\alpha$ (green) and [O III] (blue). North is at the top, and a square root stretch has been applied. The 2 arc sec. radius spectral extraction region is shown in yellow, while the sky subtraction annulus is located between the yellow and the turquoise circle, with a radius of 3 arc sec. 
  } \label{fig2}
 \end{figure}

\begin{figure}
 \centering
  \includegraphics[width =\columnwidth]{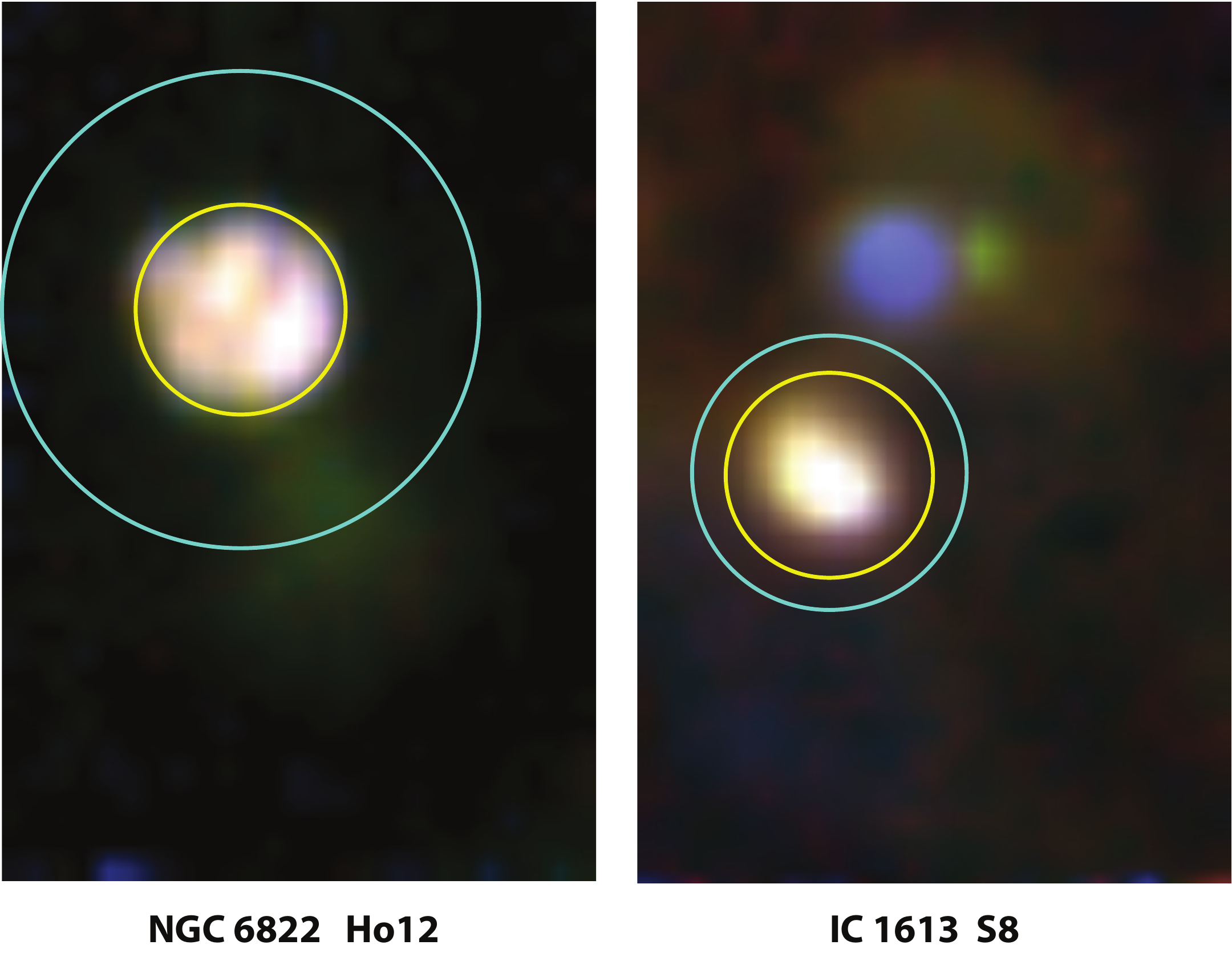}
  \caption{The extraction apertures for the SNRs NGC6822 Ho12 and IC1613 S8, shown on the WiFeS images formed from [S II] (red), H$\alpha$ (green) and [O III] (blue) using a square root stretch. For NGC6822 Ho12, north is right, while for  IC1613 S8, north is at the top. The spectral extraction region is shown in yellow, while the sky subtraction annulus is located between the yellow and the turquoise circles. Note the presence of nearby HII regions of various levels of excitation.} \label{fig3}
 \end{figure}

The spectra were extracted from the global WiFeS mosaic datacubes using using {\tt QFitsView v3.1 rev.741}\footnote{{\tt QFitsView v3.1} is a FITS file viewer using the QT widget library and was developed at the Max Planck Institute for Extraterrestrial Physics by Thomas Ott.}.  In order to extract the spectrum of the brightest shocked clouds in the resolved SNR, or to integrate over the whole SNR in the case of NGC6822 Ho12 and IC1613 S8, we used a circular extraction aperture sized to best match the size of the bright region or the full SNR. To remove the residual night sky emission and (approximately) the faint stellar contribution, we subtracted a mean sky reference annulus  surrounding the extraction aperture. The extraction regions were optimised by peaking up the signal in H$\alpha$ in the red data cube, and in H$\gamma$ in the blue data cube, respectively.  In figures \ref{fig1} to \ref{fig3} we show the extraction regions. Clearly, in the case of Kepler's SNR and the Magellanic Cloud SNR, we are subtracting the spectra of fainter parts of the SNR (whose significance is exaggerated in these images by the square-root stretch that has been used to bring out the fainter regions.

\begin{figure*}
 \centering
  \includegraphics[scale = 0.6]{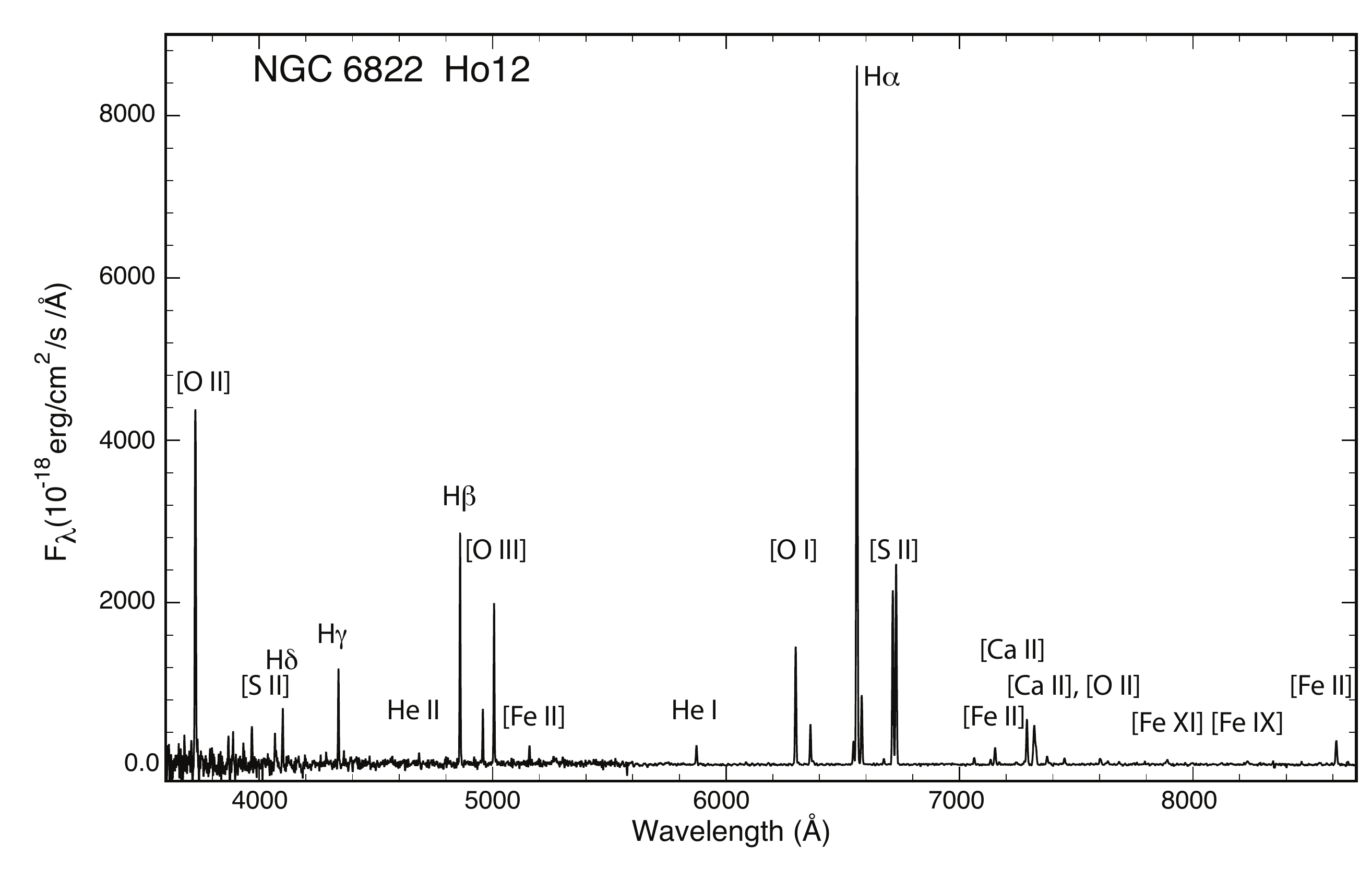}
  \caption{The extracted spectrum of NGC6822 Ho12 with some of the strongest emission lines identified.} \label{fig4}
 \end{figure*}

In figure \ref{fig4} we show the spectrum of the SNR Ho12 in NGC6822 to give an idea of the quality of the resultant extracted spectra.\newpage

\subsection{Emission Line Fluxes}

For each extracted spectrum, the spectra were first reduced to rest wavelength, and then emission-line fluxes in units of erg/cm$^2$/s, their uncertainties, the emission line FWHMs  (in \AA) and the continuum levels were measured using the interactive routines in {\tt Graf}\footnote{Graf is written by R. S. Sutherland and is available at: {\url {https://miocene.anu.edu.au/graf}} } and in {\tt Lines}\footnote{Lines is written by R. S. Sutherland and is available at: {\url {https://miocene.anu.edu.au/lines}}}. The de-reddened line fluxes relative to H$\beta$ and their uncertainties along with the measured velocity dispersions, corrected for instrumental broadening, the adopted logarithmic reddening correction at H$\beta$ ($c$), and the absolute H$\beta$ surface brightness  in units of erg/cm$^2$/s/arcsec$^2$, are given for each of the SNR in Tables \ref{tableA1} -- \ref{tableA3} in the Appendix. 

Here the reddening has been determined using a \citet{Fischera05} foreground reddening screen with a total to selective extinction $R=4.5$. This model is more realistic than a standard extinction law in extended objects, since it accounts for the fractal-like nature of the foreground attenuating dust screen, which avoids over-correction in the UV. Given that the models (described below) predict the intrinsic flux of H$\alpha$ to be in excess of three times that of H$\beta$ as a result of collisional excitation to the $n=3$ level of Hydrogen, we have iterated  the size of reddening correction applied to the observations to best match the H$\alpha$/H$\beta$/H$\gamma$/H$\delta$ ratios given by the model.

\section{Emission Line Diagnostics} \label{sec:diagnostics}

\subsection{Self-Consistent Shock Modelling}
To analyse the spectrophotometry we have built a family of radiative shocks with self-consistent pre-ionisation using the MAPPINGS 5.12 code, following the methodology described in \citet{Sutherland17}, and applied to the study of Herbig-Haro objects by \citet{Dopita17b} and to the LMC SNR N132D \citep{Dopita18}. In these models, the ionisation state of the gas entering the shock is determined by the EUV photons generated in the cooling zone of the radiative shock, and these models have an extended photoionisation zone running ahead of the shock. Although the photoionized precursor can, in principle, produce up to $\sim 25$\% of the H$\beta$ flux produced by the radiative shock, shock curvature and the finite extent of the radiative shock means that most of the ionising photons escape from the precursor region, and will instead produce an extended halo of photoionized gas around the SNR. Thus, in our previous modelling, we have ignored the contribution of the emission from the precursor.

The magnetic field pressure at the leading edge of the photo-ionised precursor (as the gas enters the shock) is assumed to be in equipartition with the gas pressure, $P_{\mathrm{mag}} = P_{\mathrm{gas}}$, and the temperature of the gas entering the shock is given by the self-consistent pre-ionisation computation; see \citet{Sutherland17} for details.

The line spectrum of a radiative shock is primarily determined by three parameters, the ram pressure driving the shock, the velocity of the shock, and the abundances of elements heavier than Hydrogen. The pre-shock magnetic field is a secondary parameter, since it serves to limit the post-shock compression factor. For slower shocks ($v_s \lesssim 120$\,km/s), the pre-shock ionisation state becomes an important factor, since partial ionisation of hydrogen entering the shock results in rapid cooling, collisional excitation of neutral hydrogen atoms into the $n=3$ level, and an enhancement in the lines of lower excitation.

Shocks faster than $\sim 180$\,km/s cool in a thermally-unstable manner \citep{Sutherland03}, in which small initial inhomogeneities are amplified during a single cooling timescale to result in a fractal distribution  of filaments with enhanced local cooling. Such instabilities are not captured in the 1-D models presented here, but may in part be responsible for the lower-velocity shocks which are a necessary element for our models to provide a good global fit to the observed spectrum.

Because these SNRs contain denser ISM clouds embedded in a less dense medium, we allow for two shock velocities driven by a common ram pressure. Typically, the fast shock will have velocity $\sim 200$\,km/s, while the slow shock will have $v_s \sim 50$\,km/s. The velocity of the slower shock is poorly constrained by the observations. The relative balance between the contributions of each of the two shock components is determined during the L1-norm minimisation procedure.   That is to say that we measure the modulus of the mean logarithmic difference in flux (relative to H$\beta$) between the model and the observations \emph{viz.};
\begin{equation}
{\rm L1} =\frac{1}{m}{\displaystyle\sum_{n=1}^{m}} \left | \log \left[ \frac{F_n({\rm model})} {F_n({\rm obs.)}} \right]  \right |. \label{L1}
\end{equation}
This procedure weights fainter lines equally with stronger lines, and is therefore more sensitive to the values of the input parameters. 

Our fitting process to the observed spectra is an iterative one. In this, we first make an initial guess of the chemical abundances, and use these to estimate the ram pressure and shock velocity by L1-norm minimisation. We then refine the abundance set to provide better agreement to the line fluxes of elements which are less important than Oxygen in determining the cooling. Next, we systematically scale the abundances of elements heavier than He over the range $\pm 0.3$\,dex in steps of 0.05\,dex to determine the best fit abundance set by L1-norm minimisation. Finally we vary the shock velocity from 100 to 300\,km/s to re-determine the best fit shock velocity by L1-norm minimisation, and make a final refinement to the abundance set. Further details of this shock fitting  procedure are presented below.

\subsection{Derivation of the Shock Parameters}
The ram-pressure driving the shocks is determined primarily using the density-sensitive line ratio [\ion{S}{2}] $\lambda\lambda 6731/6717$ (which gives the electron density in the region near the recombination zone of the shock). We also make use of the density and temperature sensitive ratios  [\ion{S}{2}] $\lambda\lambda (4067+4076)/(6717+6731)$ and  [\ion{O}{2}] $\lambda\lambda (7321+7330)/(3727+3729)$ to help constrain the ram pressure.

The abundance set is initially taken as a scaled Local Galactic Concordance (LGC)  \citep{Nicholls17}  value for each galaxy based upon the Oxygen abundances previously estimated from \HII\ region modelling, of from B-star measurements, but these are then iterated manually to achieve a better fit of the theoretical to the observed spectra once the shock velocity and ram pressure have been initially determined. In this and subsequent steps, the mixing fractions of the fast and slow shock contributions are chosen so as to minimise the value of the L1-norm.

The depletion factors of the heavy elements (caused by the condensation of these elements onto dust) are defined as the logarithm of the ratio of the gas phase abundance to the total element abundance. The  depletion factors are derived from the formulae of \citet{Jenkins09}, extended to the other elements on the basis of their condensation temperatures and/or their position on the periodic table. In these  shock models, following the findings of \citet{Dopita16} and \citet{Dopita18} we choose an initial logarithmic Fe depletion,  $\log D_{\rm Fe} = -0.5$. See \citet{Dopita18}, Table 2, for the corresponding dust depletion factors of the other refractory elements. Actual depletion factors in various ions for Fe, Ca, and Mg are determined by comparing the model predicted line fluxes with those actually observed.

\begin{figure}
 \centering
  \includegraphics[scale=0.5]{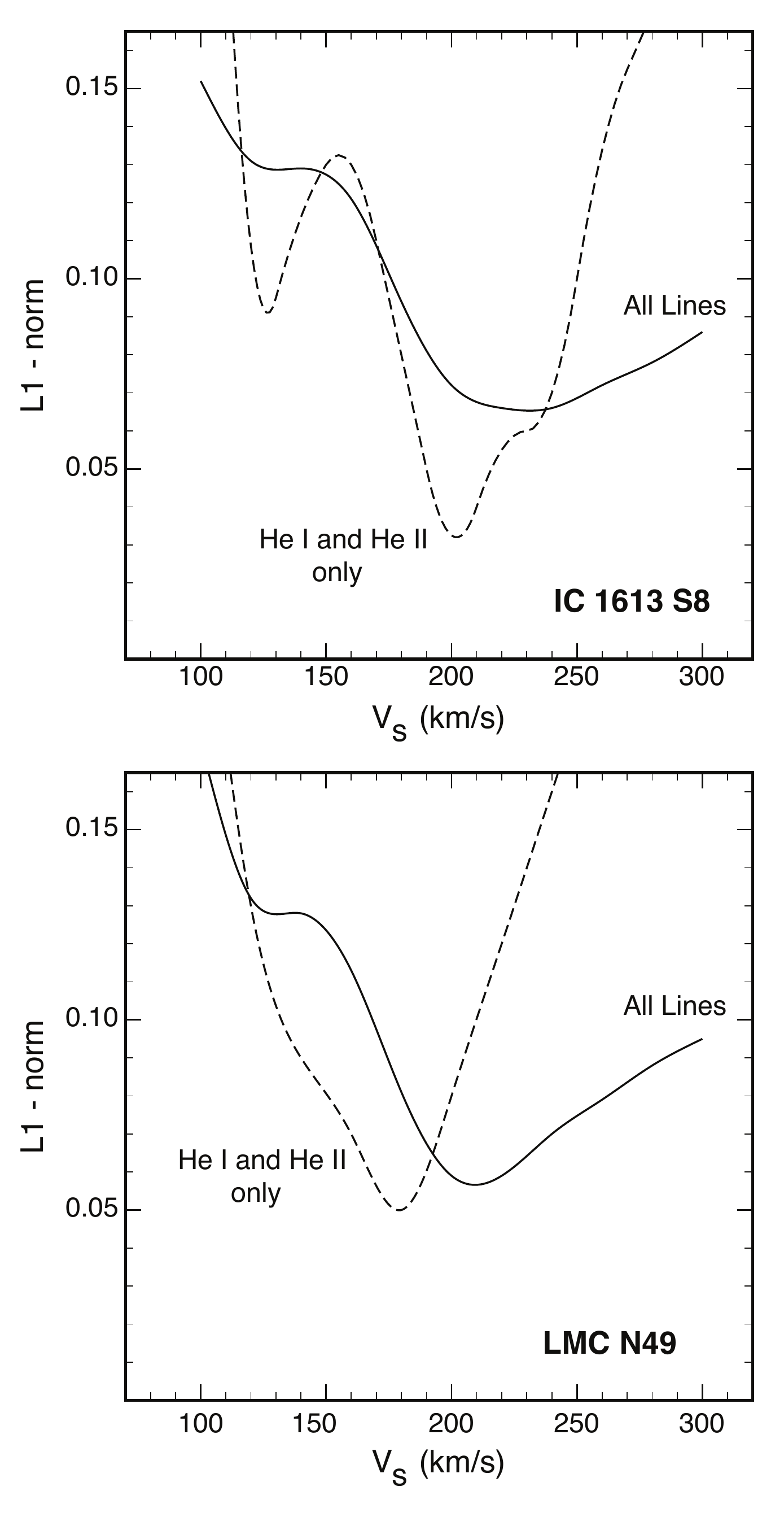}
  \caption{The velocity dependence of L1-norm of the fit to the observations measured for all lines (solid curve) and for the \ion{He}{1} $\lambda4471$ and  \ion{He}{2} $\lambda4686$ lines only (dashed curve) for  IC 1613 S8 (above) and LMC N49 (below). For both of these SNRs a shock velocity of order 200\,km/s is indicated.} \label{fig5}
 \end{figure}
\begin{figure}
 \centering
  \includegraphics[scale=0.55]{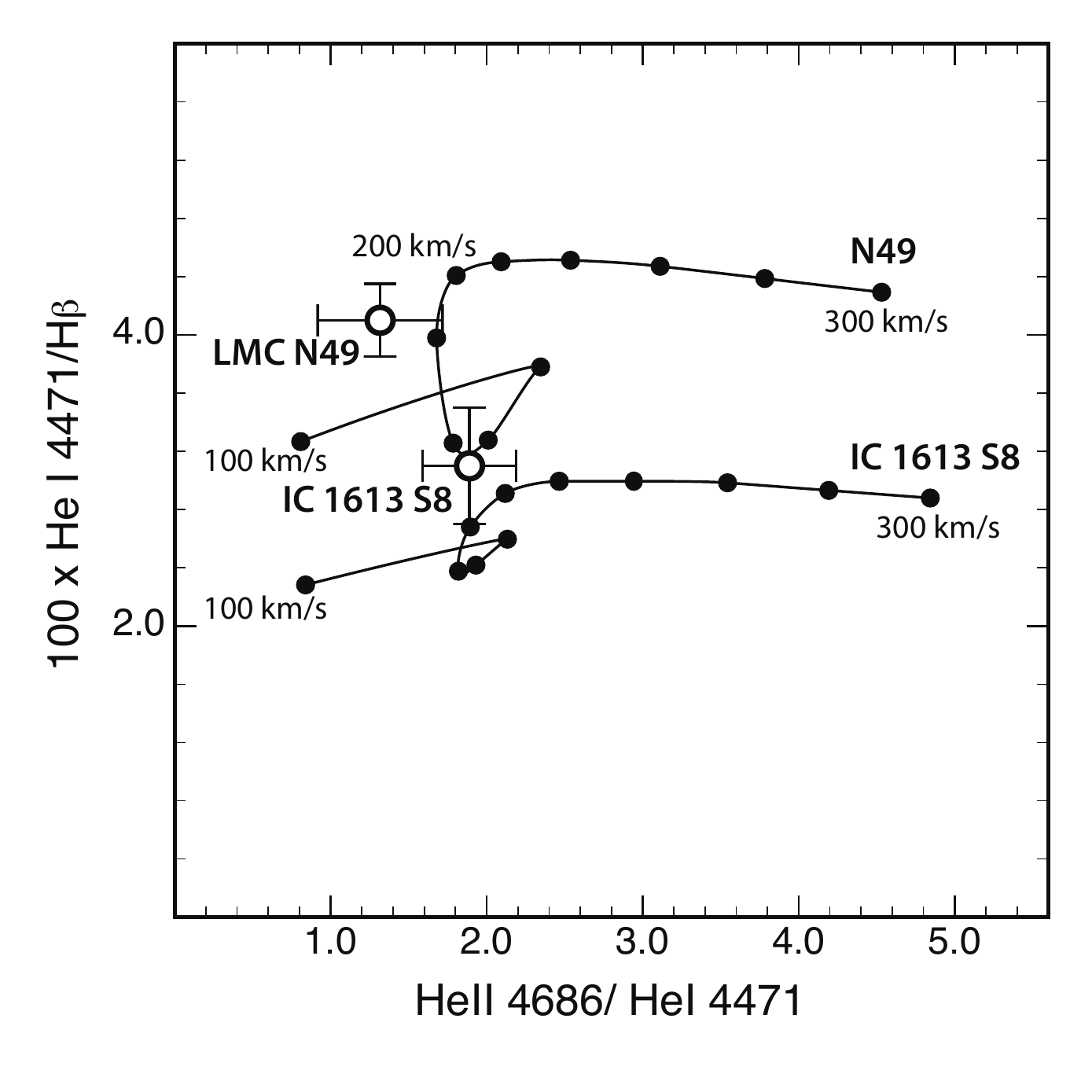}
  \caption{The velocity dependence of the \ion{He}{1}\,$\lambda4471$/H$\beta$ and the \ion{He}{2}/\ion{He}{1}\,$\lambda\lambda4686/4471$ ratio for the SNRs  IC 1613 S8 and LMC N49. The points on the theoretical lines for each galaxy are given at 20\,km/s intervals. A shock velocity of order 200\,km/s is indicated for both SNRs.} \label{fig6}
 \end{figure}

With the ram pressure fixed, we then investigate the  behaviour of the L1-norm as a function of shock velocity for the fast shock (which apart from the SMC SNR dominates the total flux). Apart from the overall behaviour of the L1-norm, this shock velocity can also be constrained by the excitation as measured by  \ion{He}{1} and  \ion{He}{2} lines. In Figures \ref{fig5} and \ref{fig6} we show two examples of L1-norm fitting, and the He line ratio fitting as a function of the shock velocity. In most of these SNR, the shock velocity in the fast shock is of order 200\,km/s, with the He lines indicating a slightly lower velocity than the global spectrum. The inferred shock velocities are fairly consistent with the measured velocity dispersions (corrected for the instrumental resolution) presented in Tables \ref{tableA1} to \ref{tableA3}, a result previously noted by \citet{Dopita12}.

 \begin{table*}
 \centering
 \small
   \caption{Observed and modelled  line intensities (scaled to H$\beta = 100$) for SNR measured with the $R=3000$ gratings.}
    \label{table2}
   \scalebox{0.7}{
   \begin{tabular}{lcrrrrrrrrrrrr}
   \\
\hline
 & Line &  \multicolumn{2}{c}{{\bf Kepler1}} &   \multicolumn{2}{c}{{\bf Kepler2}}   & \multicolumn{2}{c}{{\bf N132D}} & \multicolumn{2}{c}{{\bf SMC}}  & \multicolumn{2}{c}{{\bf NGC6822 }}  & \multicolumn{2}{c}{{\bf IC1613}}  \\
$\lambda$ \AA\ & ID & Obs. & Model & Obs. & Model& Obs. & Model& Obs. & Model& Obs. & Model& Obs. & Model\\
\hline
\\
3727,9 &  [O II] & 411.9 & 378.1 & 362.4 & 363.2 & 349.0 & 289.2 & 153.0 & 157.7 & 261.1 & 302.9 & 142.1 & 156.8\\
3870 &  [Ne III] & 75.6 & 72.7 & 75.6 & 76.7 & 32.2 & 32.3 & 4.3 & 4.6 & 16.3 & 17.9 & 9.1 & 8.7\\
3934 &  HI, Ca II  & 27.4 & 27.1 & 27.4 & 11.5 & 16.8 & 12.7 & 4.2 & 9.5 & 14.8 & 16.6 & 10.6 & 13.5\\
3967 &  [Ne III]$^{(1)}$  & 25.2 & 21.9 & 25.2 & 23.1 & 10.7 & 9.7 & 1.4 & 1.4 & 5.4 & 5.4 & 3.3 & 2.6\\
3968 &  Ca II$^{(1)}$ & 18.0 & 13.9 & 16.5 & 5.9 & 6.4 & 6.5 & 3.4 & 4.8 & 7.0 & 8.5 & 5.0 & 6.9\\
3970 &  H$\epsilon$$^{(1)}$  & 15.3 & 15.7 & 15.3 & 15.7 & 15.4 & 14.4 & 10.0 & 10.8 & 12.0 & 11.5 & 12.0 & 13.0\\
4026 &  He I  & 2.9 & 2.6 & 2.9 & 2.6 & 2.4 & 2.0 & 1.0 & 0.8 &  & 1.3 & 1.8 & 1.3\\
4069 &  [S II] & 124.2 & 101.0 & 134.2 & 142.9 & 41.4 & 39.2 & 6.5 & 5.9 & 11.0 & 13.0 & 12.6 & 14.1\\
4076 &  [S II]  & 33.2 & 32.5 & 56.8 & 46.0 & 14.1 & 12.6 & 2.9 & 1.9 & 4.4 & 4.1 & 5.1 & 4.5\\
4244 &  [Fe II]  & 7.2 & 1.7 & 6.7 & 1.6 & 4.9 & 3.2 & ... & 0.3 & ... & 0.8 & 2.6 & 1.6\\
4320 &  [Fe II]  & 4.1 & 0.5 & 1.5 & 0.4 & 0.7 & 0.3 & ... & 0.0 & ... & 0.1 & ... & 0.1\\
4340 &  H$\gamma$ & 47.7 & 46.6 & 47.5 & 46.6 & 44.8 & 45.7 & 42.6 & 42.9 & 43.3 & 43.5 & 46.1 & 45.4\\
4363 &  [O III]  & 12.9 & 13.5 & 10.6 & 13.2 & 7.1 & 9.4 & 1.2 & 3.3 & 4.8 & 6.5 & 3.1 & 4.2\\
4416 &  [Fe II]  & 11.1 & 2.1 & 14.7 & 2.0 & 4.9 & 3.6 & 1.2 & 0.5 & 6.7 & 1.7 & 3.3 & 1.8\\
4471 &  He I  & 5.5 & 5.5 & 5.0 & 5.4 & 3.9 & 3.7 & 2.4 & 1.8 & ... & 2.7 & 3.1 & 2.9\\
5467 &  Mg I]  & 8.1 & 8.1 & 10.2 & 10.2 & 3.6 & 3.6 & 1.8 & 1.8 & 2.0 & 2.0 & 1.5 & 1.5\\
4658 &  [Fe III]  & 15.3 & 9.6 & 12.8 & 10.6 & 5.1 & 4.6 & 1.4 & 1.0 & ... & 2.1 & 1.7 & 2.0\\
4686 &  He II  & 3.8 & 9.2 & 5.5 & 9.0 & 5.5 & 7.9 & 3.4 & 3.2 & 4.0 & 4.3 & 5.9 & 7.2\\
4754 &  [Fe III]   & 3.5 & 1.7 & 2.1 & 0.7 & 0.6 & 0.8 & 0.5 & 0.1 & ... & 0.4 & 0.5 & 0.4\\
4814 &  [Fe III] & 3.1 & 3.9 & 6.6 & 4.1 & 2.4 & 1.2 & 0.9 & 0.3 & ... & 0.5 & 1.7 & 0.5\\
4861 &  H$\beta$ & 100.0 & 100.0 & 100.0 & 100.0 & 100.0 & 100.0 & 100.0 & 100.0 & 100.0 & 100.0 & 100.0 & 100.0\\
4881 &  [Fe III]  & 8.5 & 10.9 & 6.2 & 12.1 & 2.1 & 4.6 & ... & 0.4 & 1.6 & 1.6 & 0.6 & 1.6\\
4949 &  [O III]  & 61.2 & 61.6 & 50.8 & 58.8 & 38.4 & 34.6 & 7.1 & 11.6 & 23.2 & 23.7 & 11.8 & 16.0\\
5007 &  [O III]  & 191.1 & 178.1 & 155.6 & 169.9 & 114.9 & 100.1 & 20.4 & 33.5 & 68.0 & 68.6 & 35.3 & 46.1\\
5016 &  He I  & 3.5 & 3.0 & 4.9 & 3.0 & ... & 2.4 & 1.0 & 1.0 & ... & 1.5 & 1.1 & 1.6\\
5158 &  [Fe II]  & 24.7 & 6.3 & 25.2 & 15.5 & 11.7 & 11.1 & 2.3 & 2.3 & 8.1 & 6.1 & 6.8 & 6.4\\
5200 &  [N I]  & 11.6 & 6.9 & 8.5 & 4.7 & 3.1 & 5.8 & 2.7 & 6.5 & 1.6 & 6.0 & 1.4 & 4.7\\
5262 &  [Fe II]  & 9.8 & 2.5 & 10.8 & 6.1 & 4.1 & 3.6 & 0.8 & 0.7 & 3.4 & 1.9 & 2.5 & 1.9\\
5271 &  [Fe II]  & 4.9 & 5.3 & 10.0 & 13.4 & 0.9 & 10.1 & ... & 0.8 & 5.3 & 4.9 & 1.9 & 4.5\\
5755 &  [N II]  & 21.5 & 9.8 & 21.7 & 7.3 & 1.3 & 2.1 & 0.2 & 0.3 & ... & 0.7 & 1.0 & 0.4\\
5876 &  He I  & 18.8 & 15.2 & 17.4 & 15.1 & 11.5 & 11.6 & 8.7 & 4.9 & 9.4 & 7.4 & 10.9 & 7.6\\
6087 &  [Fe VII]  & 2.6 & 2.6 & 1.2 & 1.2 & 0.6 & 0.6 & 0.5 & 0.5 & 1.5 & 1.5 & 1.4 & 1.4\\
6300 &  [O I]  & 144.2 & 83.7 & 149.7 & 77.5 & 104.5 & 46.6 & 49.1 & 29.7 & 58.0 & 46.2 & 38.2 & 37.1\\
6312 &  [S III]  & 2.6 & 5.7 & 2.5 & 8.4 & 0.9 & 4.0 & 0.3 & 1.2 & ... & 1.6 & 1.0 & 1.7\\
6363 &  [O I]  & 48.6 & 26.8 & 48.5 & 24.8 & 35.2 & 14.9 & 16.5 & 9.5 & 18.7 & 14.8 & 13.3 & 11.9\\
6548 &  [N II]  & 227.2 & 231.9 & 187.0 & 163.8 & 29.5 & 31.7 & 4.4 & 5.4 & 11.3 & 11.4 & 7.1 & 6.3\\
6565 &  H$\alpha$ & 308.1 & 290.5 & 301.0 & 290.4 & 305.4 & 307.2 & 342.0 & 356.0 & 338.7 & 344.1 & 331.0 & 316.2\\
6584 &  [N II]  & 692.8 & 682.2 & 494.7 & 482.0 & 91.4 & 93.2 & 16.3 & 15.8 & 33.9 & 33.5 & 20.3 & 18.6\\
6678 &  He I  & 3.6 & 4.3 & 3.5 & 4.3 & 3.2 & 3.3 & 3.2 & 1.4 & 2.7 & 2.1 & 2.8 & 2.2\\
6717 &  [S II]  & 85.5 & 84.4 & 68.8 & 76.3 & 85.8 & 81.4 & 83.4 & 111.0 & 83.8 & 89.6 & 54.5 & 86.0\\
6731 &  [S II]  & 174.5 & 178.4 & 140.6 & 161.6 & 144.9 & 138.4 & 60.8 & 78.7 & 97.2 & 109.9 & 71.1 & 109.6\\
7136 &  [Ar III]  & 10.2 & 11.5 & 10.0 & 9.4 & 4.6 & 4.8 & ... & 1.7 & 2.3 & 2.2 & 1.4 & 1.7\\
7291 &  [Ca II]  & 57.0 & 64.0 & 60.4 & 26.1 & 15.0 & 19.7 & 12.5 & 20.6 & 21.7 & 31.9 & 19.0 & 26.9\\
7319 &  [O II]$^{(1)}$  & 61.0 & 47.8 & 54.6 & 47.4 & 23.2 & 25.6 & 2.0 & 2.7 & 7.2 & 10.0 & 7.2 & 7.0\\
7323 &  [Ca II]$^{(1)}$  & 42.1 & 43.5 & 45.0 & 17.7 & 10.2 & 13.4 & 6.5 & 14.1 & 9.9 & 21.8 & 12.5 & 18.4\\
7329 &  [O II]  & 49.7 & 38.9 & 46.0 & 38.5 & 18.8 & 20.7 & 1.4 & 2.2 & 5.9 & 8.1 & 5.0 & 5.6\\
7637 &  [Fe II]  & 6.6 & 2.9 & 6.6 & 3.1 & 4.3 & 1.7 & 1.5 & 0.4 & 1.5 & 1.1 & 1.1 & 1.1\\
7751 &  [Ar III]  & 3.5 & 2.8 & 2.2 & 2.3 & ... & 1.2 & ...  & 0.4 & ... & 0.5 & ... & 0.3\\
8579 &  [Cl II]  & 5.2 & 3.7 & 3.8 & 3.4 & 1.8 & 2.0 & 0.4 & 0.4 & ... & 0.7 & ... & 1.5\\
8617 &  [Fe II]  & 46.2 & 32.1 & 41.8 & 34.3 & 27.9 & 19.4 & 3.6 & 4.8 & 12.5 & 12.3 & 9.7 & 12.7\\
8727 &  [C I]  & 3.2 & 3.2 & 3.2 & 3.2 & 1.2 & 1.3 & 1.0 & 1.0 & ... & 0.6 & ... & 0.3\\
8891 &  [Fe II]  & 13.1 & 10.9 & 14.5 & 11.7 & 7.5 & 6.1 & 0.8 & 1.6 & 3.8 & 3.8 & 3.1 & 3.9\\
&&&&&&&&&&&&& \\
\hline
&&&&&&&&&&&&& \\
 \multicolumn{13}{l}{\footnotesize{$^{1}$ These blends have been separated with the help of the relative fluxes given by the model.}} 
  \end{tabular}}
\end{table*}

\subsection{Modelling Individual Objects}
Given that, apart from a wide range in chemical abundances, the pre-shock densities, evolutionary status and the supernova type of these SNR differ widely, each present different modelling challenges. Here we summarise the main issues.
\newline
\newline
{\bf Kepler's SNR~~} This remnant of the historical supernova of SN 1604 was Type Ia  \citep{Baade43, Yamaguchi14}. Recently \citet{Sankrit16} have made a detailed proper-motion study of the blast wave using \emph{Hubble Space Telescope} images separated by about 10\,yr to establish a distance of $5.1^{+0.8}_{-0.7}$\,kpc to the remnant, which with Galactic coordinates G4.5+6.8 puts it almost directly towards the Galactic centre. The spectrophotometry of the bright shocked clouds by \citet{Dennefeld82} and \citet{Leibowitz83} showed that these were very dense ($n_e \sim 1000$\,cm$^{-3}$), which led to the suggestion that they represent shocked circum-stellar material, rather than pristine samples of the ISM. However, densities this high are encountered in other SNR, notably in the LMC SNR N132D \citep{Dopita18} and these are the consequence of the blast wave encountering the dense self-gravitating clouds of the ISM. { We discuss this possibility further below.}

Blast wave velocities of up to $\sim 3000$\,km/s were inferred by \citet{Sankrit16}, making the radiative shocks in the dense clouds very young. Indeed, one of these, the so-called ejecta knot first appeared only around 1970 \citep{vandenBergh77}. Given their positions relative to the blast wave shell, the bright regions observed here (the Box1 region of \citet{Sankrit16}) cannot be much older. We therefore model them as finite-age shocks with radiative ages of $\sim 50$\,yr.\newline
\newline
{\bf N49 \& N103B~~} The difficulty in determining abundances in these objects is the fact that the $R=7000$ grating was used in the red, meaning that the spectra between $\lambda\lambda 7100 - 9000$ are not observed. This precludes a determination of either the C abundance using the [\ion{C}{1}] 8727\AA\ line, the Ar abundance using the [\ion{Ar}{3}] 7135\AA\ or 7751\AA\ lines, and the Cl abundance using the [\ion{Cl}{2}] 8579\AA\ line. Furthermore, the measurement of the dust depletion factors is compromised by the lack of a measurement of the IR lines of [\ion{Fe}{2}], [\ion{Ca}{2}] and [\ion{Ni}{2}].
\newline
\newline
{\bf SMC SNR 0104-72.3~~} This SNR is rather faint, and our measurements are of an extensive region of somewhat enhanced density towards the eastern boundary of the SNR centred at 01:06:24.0 -72:05:35 (J2000). This region is dominated by the slow shock component, which renders the determination of the fast shock parameters more uncertain. The faster shocks are marked by the (blue) [\ion{O}{3}] emission  in Figure \ref{fig1}. As can be seen this emission tends to delineate the current position of the blast wave.
\newline
\newline
{\bf NGC6822 Ho12 \& IC1613 S8~~} In both of these cases, we are determining the global spectrum of the SNR. If there are variations in the local shock conditions, which is certainly the case \citep{Kong04}, these will be lost in the averaging process.

\subsection{Fits to Observed Spectra}
\begin{figure*}
 \centering
  \includegraphics[scale =0.65]{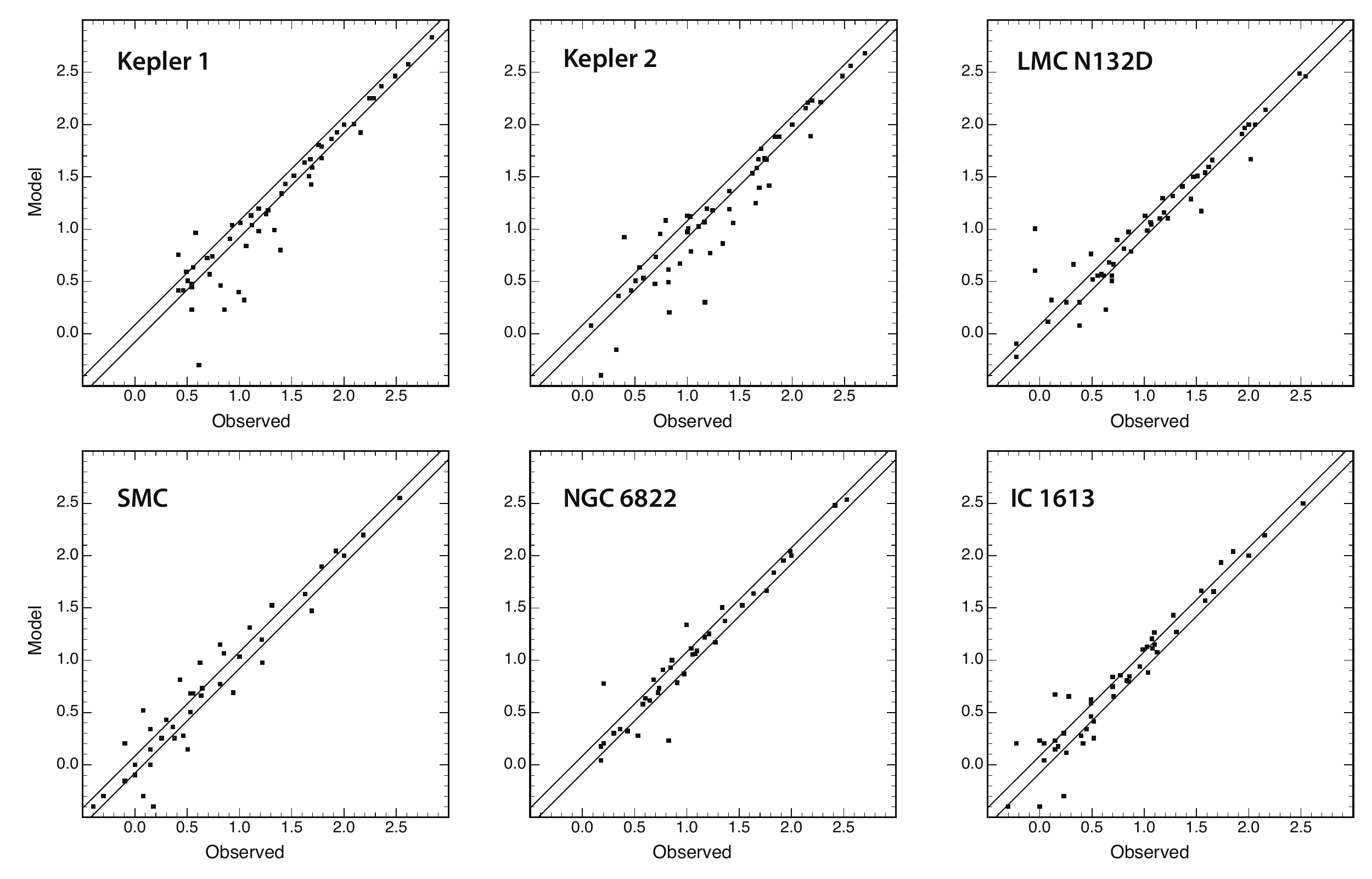}
  \caption{A comparison on the observed and modelled line fluxes plotted on a logarithmic (Base 10) scale with $\log F_{H\beta} =2.0$. The solid lines represent a difference between the model and the observation of $\pm20$\%.} \label{fig7}
 \end{figure*}

While the plot of the L1-norms is useful to summarise the quality of the fit, it is instructive to see how the individual line intensities predicted by the model fit to the observations. We show this in Table \ref{table2}, which gives the quality of the fits to over 50 lines for those objects observed with the $R=3000$ gratings. The fits are depicted graphically in Figure \ref{fig7}. Here, we have used the relative strengths of lines in blends as predicted by the model to disentangle the individual line fluxes from Table \ref{tableA1} -- \ref{tableA3}. We have corrected the line fluxes given by the model for [\ion{Fe}{2}], [\ion{Fe}{3}], [\ion{Fe}{7}], [\ion{Mg}{1}], \ion{Ca}{2} and [\ion{Ca}{2}] using the depletion factors derived in Table \ref{table3}, below. 

Overall the fit to the models is excellent. The outstanding issues apparent from Table \ref{table2} are as follows. First, the models tend to over-estimate the strength of the  [\ion{N}{1}] $\lambda\lambda 5198,5200$ doublet, while under-estimating the strength of the  [\ion{O}{1}] $\lambda\lambda 6300, 6363$ doublet. Second, the models do not quite succeed in reproducing  the \ion{He}{1}\,$\lambda4471$/H$\beta$ and the \ion{He}{2}/\ion{He}{1}\,$\lambda\lambda4686/4471$ ratios, as can be seen in Figure \ref{fig6}. For ease of comparison, we present the comparison of the modelled and observed line fluxes for all the fitted lines in Figure \ref{fig7}.

{ The [\ion{N}{1}] and  [\ion{O}{1}]  lines are formed in the cool, partially-ionised region deep within the recombination zone of the shock region which is particularly difficult to model. The predicted line strengths are strongly dependent on the ionisation parameter of the local hard radiation field, which is in turn strongly dependent on the detailed geometry of the recombination zone. This geometry also depends on the degree of thermally-unstable cooling, which is capable of generating many local low-velocity shocks with a complex recombination zone geometry \citep{Sutherland03}. Such issues will also affect the predictions of the [\ion{C}{1}] line and the derived C abundance. However, it is hard to properly quantify these errors. We have placed an extra $\pm0.15$\,dex error on the C abundance to make allowance for this uncertainty.}

\section{Abundances and Depletion Factors}\label{abund}
\subsection{Abundances}
\begin{figure}
 \centering
  \includegraphics[scale=0.6]{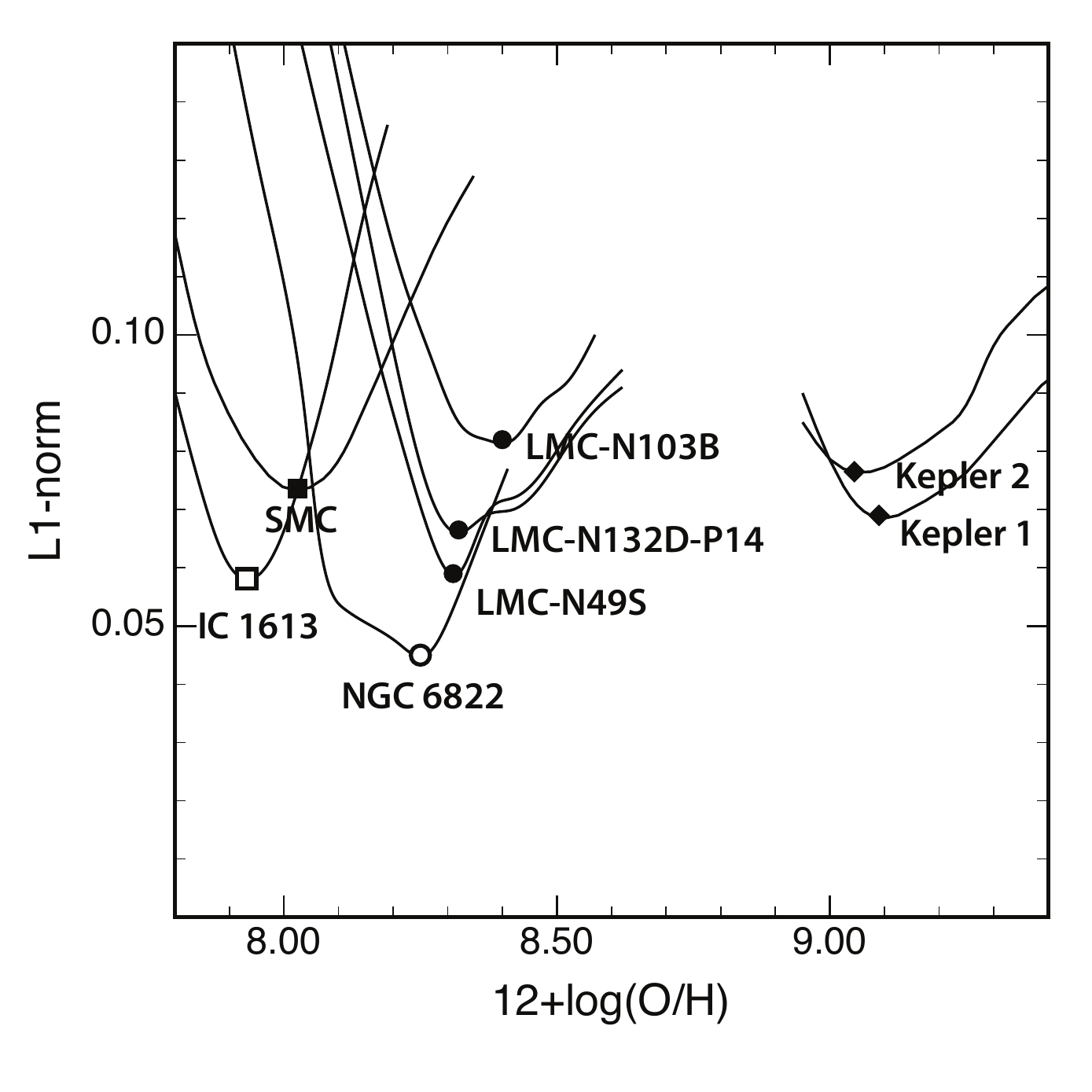}
  \caption{The behaviour of the L1-norm as a function of the scaled O abundance for the observed SNRs. The shock velocities, ram pressures and mixing ratio of the fast and slow shock components are kept fixed in these models. The best-fit abundances of each galaxy are marked.} \label{fig8}
 \end{figure}

In Figure \ref{fig8} we present a summary of the L1-norm abundance fitting for each of the observed SNR. In each case a clear minimum is found, although this tends to be shallower for the highest abundance objects. This is due to ``saturation'' of the emission line intensities, as each of the principal cooling ions compete for the limited amount of enthalpy in the cooling zone in which they emit. However, it is clear that these SNR measurements constrain the global O/H abundance to within $\pm0.1$\,dex.

In Table \ref{table3} we list the derived shock parameters -- the ram pressure driving the shocks, $P_{\mathrm{ram}}$, the velocity of the fast shock, $V_{\mathrm{s}}$, and the fraction of the H$\beta$ flux emitted by the fast shock, $F$. In the case of Kepler's SNR, the shocks are of finite radiative age (see below). We also give the estimated age of these in Table  \ref{table3}. 

The derived chemical abundances and their estimated errors are also given in the table. For O the errors can be gauged from Figure \ref{fig8}. These are determined primarily from the [\ion{O}{2}] and [\ion{O}{3}] fitting. The predictions of the models for the [\ion{O}{1}]  lines are less secure, since these arise in a narrow region around the H recombination zone. { As discussed above}, our models tend to systematically under-estimate the strength of the [\ion{O}{1}] lines while over-estimating the strength of the [\ion{N}{1}] lines, which arise in the same zone. The errors in determining the He and the O abundances are appreciably less than for those of other elements, since most of these depend on the measured flux in a single line or doublet. The derived abundances of N and S are of intermediate accuracy, since these have good measurements of both their blue and red lines. Estimating the errors in any particular measurement of the elemental abundance is difficult, given the non-linear nature of the fitting procedure. However, a further indicator of the errors can be derived from the scatter in the abundances derived from the two regions in the Kepler's SNR, and from the scatter in the abundances derived for the three LMC SNRs.

 \begin{table*}
 \centering
 \small
   \caption{Derived Shock Parameters, Abundances ($12+ \log X/H$), and inferred ionic depletion factors.}
    \label{table3}
   \scalebox{0.7}{
  \begin{tabular}{lcccccccc}
  \\
 \hline
 & {\bf Galaxy} & {\bf Galaxy}& {\bf LMC}  &  {\bf LMC}   &  {\bf LMC}   &  {\bf SMC}  & {\bf NGC6822} &{\bf IC1613} \\
 & Kepler 1 & Kepler 2 & N49 & N103B & N132D & 0104-72.3 & Ho 12 & S8\\
 \hline
 &  &  &  &  &  &  &  & \\
{\bf Shock} &  &  &  &  &  &  &  & \\
{\bf Parameters:}. &  &  &  &  &  &  &  & \\
$P_{\mathrm{ram}}$ (dynes/cm$^2$) & 1.81E-07 & 1.80E-07 & 2.67E-07 & 1.33E-06 & 3.52E-07 & 1.82E-09 & 6.19E-08 & 3.70E-08\\
$V_{\mathrm{s}}$ (km/s) & 220 & 220 & 210 & 220 & 200 & 230 & 200 & 230\\
Fast Shock Fract. $F$ & 1.00 &1.00 & 0.82 & 0.79 & 0.88 & 0.37 & 0.53 & 0.60\\
Shock Age (yr) & 50 & 40 &  &  &  &  &  & \\
\hline
&  &  &  &  &  &  &  & \\
{\bf Elemental} &  &  &  &  &  &  &  & \\
{\bf Abundance:} &  &  &  &  &  &  &  & \\
He & $11.02\pm0.03$ & $11.06\pm0.03$ & $10.96\pm0.03$ & $10.92\pm0.03$ & $10.96\pm0.03$ & $10.91\pm0.03$ & $10.95\pm0.03$ & $10.94\pm0.03$\\
C & $9.06\pm0.32$ & $8.99\pm0.32$ & $8.09\pm0.32$ & (8.09) & $8.09\pm0.32$ & $7.50\pm0.32$ & (7.45) & $7.20\pm0.32$\\
N & $8.44\pm0.08$ & $8.27\pm0.08$ & $7.17\pm0.05$ & $7.17\pm0.05$ & $7.32\pm0.05$ & $6.82\pm0.06$ & $6.99\pm0.07$ & $6.67\pm0.05$\\
O & $9.09\pm0.06$ & $9.04\pm0.06$ & $8.32\pm0.04$ & $8.37\pm0.04$ & $8.32\pm0.04$ & $8.02\pm0.06$ & $8.25\pm0.06$ & $7.92\pm0.04$\\
Ne & $8.49\pm0.10$ & $8.32\pm0.10$ & $7.66\pm0.05$ & $7.57\pm0.05$ & $7.62\pm0.04$ & $7.04\pm0.10$ & $7.52\pm0.06$ & $7.01\pm0.05$\\
Mg & (7.70) & (7.65) & (7.19) & (7.19) & (7.19) & (6.64) & (6.95) & (6.52)\\
Si & (7.64) & (7.59) & (7.11) & (7.11) & (7.11) & (6.77) & (6.89) & (6.66)\\
S & $7.84\pm0.07$ & $7.79\pm0.07$ & $7.03\pm0.09$ & $6.88\pm0.09$ & $7.10\pm0.07$ & $6.79\pm0.13$ & $6.84\pm0.07$ & $6.69\pm0.06$\\
Cl & $5.68\pm0.12$ & $5.58\pm0.12$ & (4.96) & (4.96) & $4.96\pm0.10$ & $4.37\pm0.15$ & (4.49) & (4.04)\\
Ar & $6.43\pm0.14$ & $6.23\pm0.14$ & (5.79) & (5.79) & $5.79\pm0.10$ & (5.49) &$5.65\pm0.14$ & $5.33\pm0.14$\\
Ca & (6.46) & (6.46) & (6.02) & (6.02) & (6.02) & (5.67) & (5.91) & (5.28)\\
Fe & (7.66) & (7.66) & (7.33) & (7.33) & (7.33) & (6.76) & (7.13) & (6.75)\\
Ni & (6.34) & (6.34) & (5.91) & (5.91) & (5.91) & (5.55) & (5.79) & (5.36)\\
 &  &  &  &  &  &  &  & \\
 \hline
{\bf Depletion:} &  &  &  &  &  &  &  & \\
{\bf $\log D$ } &  &  &  &  &  &  &  & \\
Fe II & -0.15 & -0.08 & -0.39 & -0.47 & -0.24 & -0.22 & -0.22 & -0.11\\
Fe III  & -0.14 & -0.08 & -0.80 & -0.76 & -0.61 & -0.36 & -0.64 & -0.41\\
Fe VII  & -0.19 & -0.49 & -1.11 & -0.87 & -0.98 & -0.39 & -0.24 & -0.16\\
 &  &  &  &  &  &  &  & \\
Ca II & 0.12 & 0.14 & -0.32 & -0.19 & -0.35 & -0.28 & -0.10 & 0.20\\
Ni II  & 0.09 & 0.09 & ... & ...  & ... & -0.20 & -0.22 & ...\\
Mg I & -0.25 & -0.13 & -0.59 & -0.46 & -0.49 & -0.29 & -0.55 & -0.32\\
&  &  &  &  &  &  &  & \\
 \hline
 \multicolumn{9}{l}{\footnotesize{Abundance values given in parentheses are the values assumed in the model.}} 
  \end{tabular}}
\end{table*}

\subsection{Depletion Factors}
{ The derived depletion factors for the refractory elements depend on what is assumed for the Ca, Mg, Fe and Ni abundances. For these we use the Ca/O, Mg/O and Fe/O scaling vs. O/H abundance, as given by \citet{Nicholls17}. For Ni, we assume that the Ni/O ratio scales as the Fe/O ratio.}

All the models were run with a fixed pattern of depletions of the refractory elements from the gas onto dust, { using the \citet{Jenkins09} scaling relations for an iron depletion factor of -0.5}. This gives  $\log D_{\rm Fe} = -0.50$, $\log D_{\rm Ca} = -0.36$, $\log D_{\rm Mg} = -0.00$ and $\log D_{\rm Ni} = -0.40$. The ratios of the observed to predicted line intensities different elements and ionisation stages were then compared to derive the actual value of the depletion factor required to best match the observations, resulting in the derived depletion factors listed in Table \ref{table3}.

{ Note that, in Table \ref{table2} we do not attempt to fit the lines of the higher excitation species of iron, such as [\ion{Fe}{10}] and [\ion{Fe}{14}]. These species arise in the faster non-radiative shocks located as bow shocks around the dense clouds measured here, as was clearly demonstrated in \citet{Dopita16,Dopita18} for the LMC SNRs, N49 and N132D.}

{ Within the errors, the depletion factors derived for Kepler's SNR are zero, consistent with almost total destruction of the dust grains. This result is not inconsistent with the IR IRAC and MIPS observations by \citet{Blair07}. These authors found that the current mass of dust in the warm dust component is $5.4\times10^{-4}$\,M$_{\odot}$, and they inferred an original dust mass of about $3\times10^{-3}$\,M$_{\odot}$ before grain sputtering. These figures would imply $\log D \sim -0.09$, which is similar to the [\ion{Fe}{2}] and [\ion{Fe}{3}] depletion factors derived here. This high dust destruction degree is presumably related to the high density in the pre-shock gas of the fast shock ($n_H \sim 600$\,cm$^{-3}$).} 

For the remaining SNRs, which are all likely to result from Type II SNe, $ -0.47 < \log D_{\rm FeII} <  -0.11$,  $ -0.80 < \log D_{\rm FeIII} < -0.36$ and  $ -1.11 < \log D_{\rm FeVII} < -0.24$. This result is consistent with the conclusions of \citep{Dopita16} and \citep{Dopita18} that dust has been mostly destroyed in the region emitting the [\ion{Fe}{2}] lines, while a smaller fraction has been destroyed in the \ion{Fe}{3} and \ion{Fe}{7} zones, consistent with the gyro-spinup grain destruction models of \citet{Seab83} and \citet{Borkowski95}.

\section{Discussion}\label{discuss}
\subsection{Shock Velocities}
{ From Table \ref{table3}, it is apparent that we derive a fast shock velocity in the range $200-230$\,km/s for all the observed SNRs. This seems to be an extraordinarily narrow range which demands some physical explanation.

It is certainly true that our models with either a single or two shock velocities are an over-simplification of reality, and that a wider range in shock velocities must be present. If we consider an ensemble of shocks driven into a medium with variable density by a common driving pressure, $P$, then the shock luminosity $L$ of a shock with velocity $v$ is given by  $L \propto \rho v^3$ or, equivalently, $L \propto Pv$. Therefore the most luminous shock will be the fastest fully radiative shock. At a shock velocity of  $200-230$\,km/s, the plasma is heated to the peak of the cooling function, so in general this velocity will correspond to the fastest radiative shock.}

\subsection{Kepler's SNR}
Returning to the question of whether the shocked clouds in Kepler represent circum-stellar or true interstellar matter, we note that the estimated $12 +\log(\mathrm{\mathrm{O/H}})$ of these clouds is $9.06\pm0.1$. If the difference between this and the Local Galactic value, $8.72$, is solely due to a galactic abundance gradient, then at a distance of  $5.1^{+0.8}_{-0.7}$\,kpc derived by \citet{Sankrit16}, the Galactic Abundance gradient in O would be $-0.059\pm 0.022$\,dex/kpc. This is in close agreement with that determined by \citet{Rolleston00} from 80 early B-type stars, $-0.067\pm0.008$\,dex/kpc and with that found from the study of H II regions as cited by \citet{Rolleston00}.  

An extensive review by \citet{Gillessen13} concludes that the most likely distance to the Galactic centre is $8.20\pm0.35$\,kpc. Using this distance puts Kepler's SNR at a distance of $\sim3.1$\,kpc from the Galactic centre. From the work of \citet{Smartt01} who studied the abundances of stars within the inner 5\,kpc of the Galaxy, we expect  $12 +\log(\mathrm{O/H}) = 9.0\pm0.2$, $12 +\log(\mathrm{N/H}) = 8.3\pm0.2$ at $R=3.1$\,kpc, while we measure Kepler's SNR to have $12 +\log(\mathrm{O/H}) = 9.06\pm0.10$ and $12 +\log(\mathrm{N/H}) = 8.35\pm0.10$. 

The high value of the derived C abundance ( $12 +\log(\mathrm{C/H}) = 8.99-9.06$) might also be taken as an indication that the clouds in Kepler are of circum-stellar origin. However, at high abundances, C behaves as a secondary nucleosynthesis element such that the C/O ratio increases in proportion to the O abundance. The abundance scaling given in \citet{Nicholls17} suggests that, at the O abundance measured in Kepler, we would expect to see $\log(\mathrm{C/O)} = -0.05\pm0.10$, which is entirely consistent with the observed value. Thus we conclude that the shocked clouds in Kepler's SNR are truly of interstellar composition, given their location within the Galaxy.

{ It is clear, however, that the Kepler clouds are not representative of a diffuse component of the ISM, given that the inferred pre-shock density of these clouds is $n_{\mathrm{H}} \sim 600$\,cm$^{-3}$.  Such high densities are atypical at a distance of $\sim 500$\,pc from the Galactic plane and are also inconsistent with  hydrodynamical simulations that indicate a medium with some symmetry about the SN site \citep{Tsebrenko13}.

It is more likely that the ``Box 1" cloud complex of  \citet{Sankrit16} represents a self-gravitating entity. Using our measured pre-shock density and adopting a diameter of $\sim30-40$\,arc sec derived from the IR image presented in \citet{Sankrit16}, we infer a cloud mass of $1.4-3.5$\,M$_{\odot}$.  This value is comparable to the masses derived by \citet{Dopita18} for the shocked clouds in the LMC supernova remnant N132D;  $0.1-20$\,M$_{\odot}$ with a mean of $\sim 4$\,M$_{\odot}$. These N132D clouds were inferred to represent the cool ISM and to be typical ISM self-gravitating Bonnor-Ebert spheres such as those recently investigated on a theoretical basis by \citet{Sipila11, Sipila17} and \citet{Fischera14}. The nature of the Kepler clouds is probably similar.

Because the shocked clouds in Kepler appear to represent gravitationally-confined samples of the ISM as they existed before the supernova event, and because these clouds are dense enough that they would not be appreciably affected in their chemical composition by any pre-supernova mass-loss, they should therefore present ideal samples of pristine ISM in the galactic regions close to the Galactic Center.}

\subsection{Abundance Scaling Relations}
\subsubsection{Oxygen}
In Table \ref{table4}, we present the comparison of the derived SNR O-abundances for these SNRs, and for stars and \HII\ regions. For the \HII\ regions, we distinguish determination made using the collisionally-excited lines (CELs) and using recombination lines (RLs). In the case of Kepler, we use the stellar abundances from \citet{Smartt01} and for the \HII\ regions the data from Afflerbach et al. (1997). For consistency, the stellar and \HII\ region measurements of the remaining galaxies, we have used the curated compilation of \citet{Bresolin16} (for the data sources they used see their Table 3).

Overall, the O abundances determined using SNRs agree with the stellar abundance scale to 0.034\,dex, the \HII\ region CEL abundance scale to 0.024\,dex, and to the mean of the stellar and \HII\ regions to 0.034\,dex. It is clear that the RL abundance scale lies systematically high relative to the other methods of determination. The most likely cause of this offset is fluorescent excitation of lines of Oxygen by the UV radiation field of the central stars, as quantified in the study of the planetary nebula IC\,418 by \citet{Morisset09}. This object is relevant to the study of \HII\ regions since the central star has a vey similar effective temperature to the O-stars which excite \HII\ regions.

 A second source of the discrepancy could be the existence of a $\kappa-$distribution in the electrons \citep{Nicholls12, Dopita13}. Recently, \citet{Livadiotis18} has provided the theoretical underpinning of this hypothesis by proving that the most general, physically meaningful, distribution function that particle systems are stabilized into when reaching thermal equilibrium is the kappa distribution family, of which the Maxwell-Boltzmann distribution is a limiting case.

 \begin{table}
 \centering
 \small
   \caption{Oxygen abundance scaling}
    \label{table4}
     \scalebox{0.9}{
      \begin{tabular}{lcccc}
      \\
      \hline
    {\bf Galaxy} & {\bf SNRs} & {\bf Stars} & \multicolumn{2}{c}{{\bf \HII\ Regions}} \\
    & & & CELs & RLs \\
\hline   
Galaxy & & & & \\ 
 (\emph{R=3.1kpc}) & $9.06\pm0.08$ & $9.00\pm0.20$ & $9.10\pm0.20$ & ..\\
LMC & $8.34\pm0.05$ & $8.33\pm0.08$ & $8.40\pm0.10$ & $8.54\pm0.05$\\
SMC & $8.02\pm0.06$ & $8.06\pm0.10$ & $8.05\pm0.09$ & $8.24\pm0.16$\\
NGC\,6822 & $8.25\pm0.06$ & $8.08\pm0.21$ & $8.14\pm0.08$ & $8.37\pm0.09$\\
IC\,1613 & $7.92\pm0.04$ & $7.90\pm0.08$ & $7.78\pm0.07$ & ... \\
 
  & & & & \\
 \hline
  \end{tabular}}
\end{table}

\subsubsection{Nitrogen}
In determining the interstellar medium (ISM) abundances of galaxies from their \ion{H}{2} regions, we are heavily reliant on an accurate measurement of the N/O ratio, since this ratio increases continuously with O/H as N starts to behave as if it were a secondary nucleosynthetic element. For example, one of the best abundance diagnostics is based upon the [\ion{N}{2}]/[\ion{O}{2}] ratio \citep{Kewley02}. At high redshift the [\ion{O}{2}]$\lambda\lambda3727,3729$ doublet is often unobservable, and calibrations based upon the  [\ion{N}{2}]/H$\alpha$ ratio \citep{Denicolo02}, the  [\ion{N}{2}]/[\ion{O}{3}] ratio \citep{Pettini04} or else using the [\ion{N}{2}], H$\alpha$ and [\ion{S}{2}]  together \citep{Dopita16b} are used instead. 

\begin{figure}
 \centering
  \includegraphics[scale=0.6]{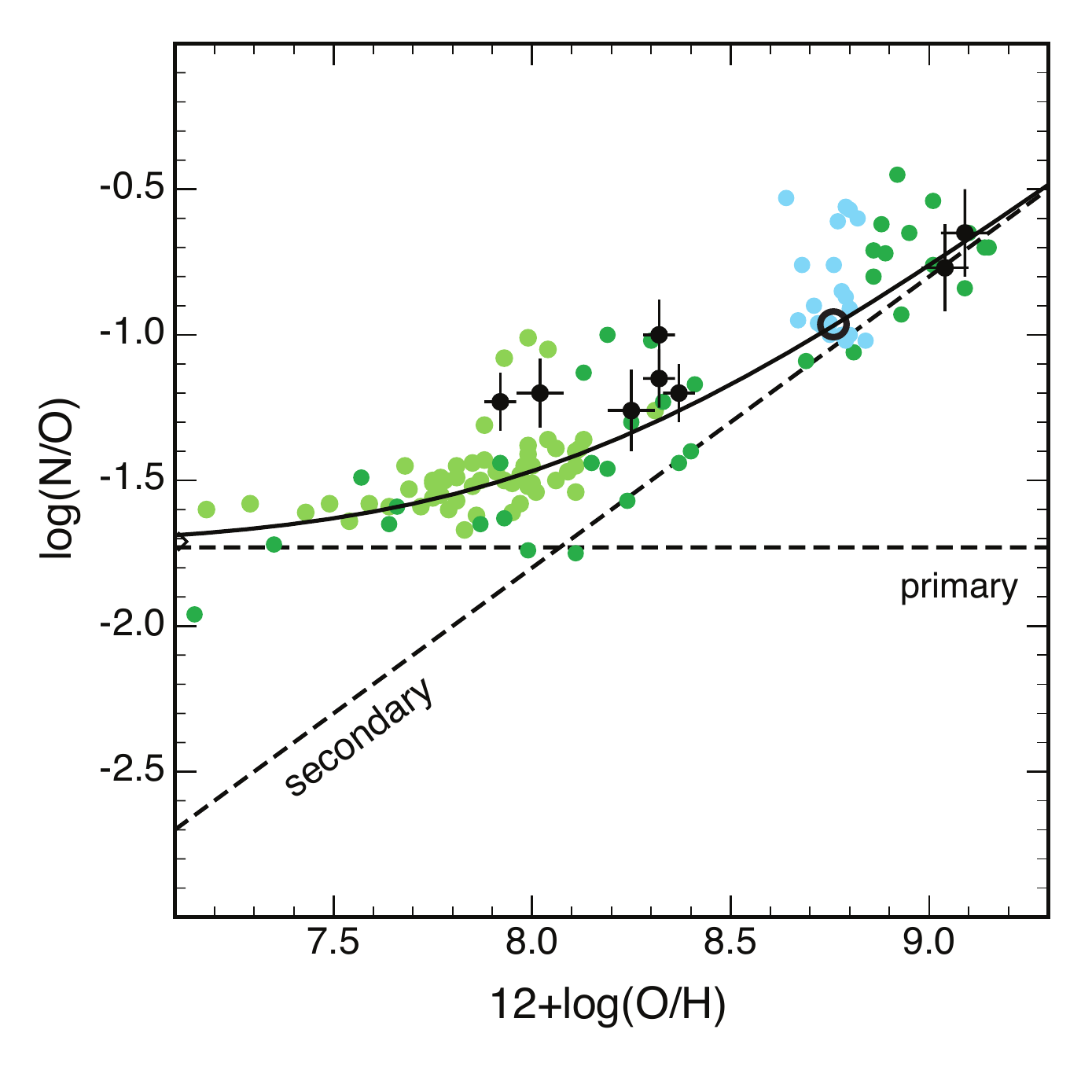}
  \caption{ $12 +\log(\mathrm{O/H})$ vs. $\log(\mathrm{N/O})$ for our SNRs (black filled circles), the Sun (black open circle), and the stellar and \HII\ region determinations referenced in the text (green and blue circles). The solid black curve is the best-fit relation presented by \citet{Nicholls17} by adopting primary and secondary nucleosynthetic components shown as dotted lines.} \label{fig9}
 \end{figure}

In Figure \ref{fig9} we compare our SNR results with other works on the  $12 +\log(\mathrm{O/H})$ vs. $\log(\mathrm{N/O})$ diagram. Following \citet{Nicholls17}, we have drawn our data from \citet{Spite05} which refer to halo metal-poor unmixed giants, \citet{Israelian04} who made UV observations of unevolved metal-poor stars, \citet{Fabbian09} who measured halo solar-type dwarfs and sub-giants,\citet{Nieva12} who observed local B stars,  and with nebular data collected from Blue Compact Galaxies from \citet{Izotov99}.  Clearly the agreement between the SNR, \HII\ region and stellar measurements are within the errors, suggesting that the $12 +\log(\mathrm{O/H})$ vs. $\log(\mathrm{N/O})$  calibration in the Local Universe is now well-determined.

\subsubsection{Ne, Ar and Cl}
\begin{figure}
 \centering
  \includegraphics[scale=0.55]{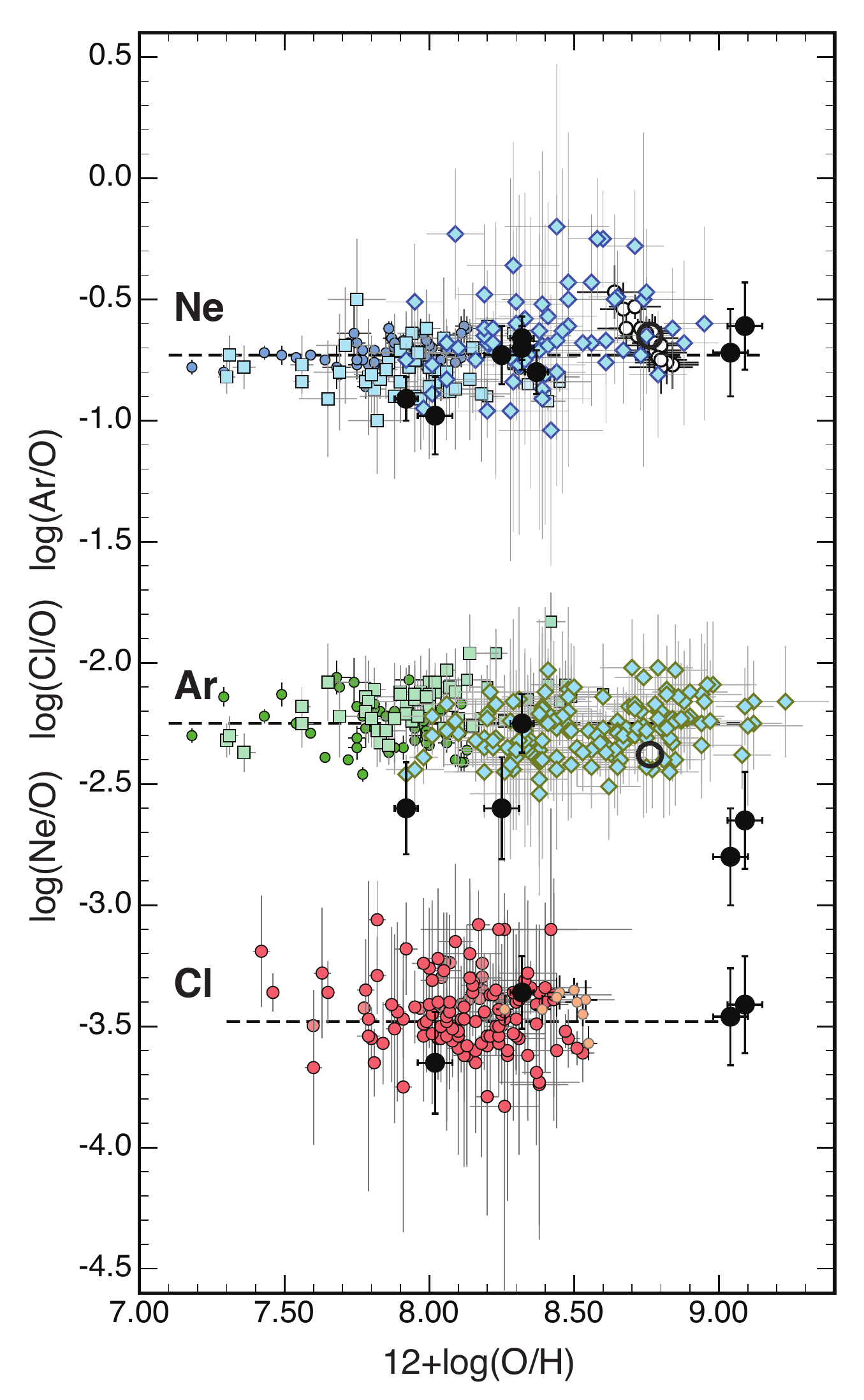}
  \caption{ $12 +\log(\mathrm{O/H})$ vs. $\log(\mathrm{Ne/O})$,  $\log(\mathrm{Ar/O})$ and  $\log(\mathrm{Cl/O})$ for our SNRs (black filled circles), the Sun (black open circle), and the stellar and \HII\ region determinations referenced in the text. For Ne and Cl the scaling found for the SNRs agrees with the other determinations.} \label{fig10}
 \end{figure}

Neon and Argon are both $\alpha-$process elements (and should therefore scale as Oxygen), and are noble gases (and therefore suffer no depletion onto dust grains). Following \citet{Nicholls17}, in Figure \ref{fig10} we plot the measured ratio of these elements with respect to Oxygen vs. $12 +\log(\mathrm{O/H})$ from our SNRs, and from both stellar and nebular sources taken from \citet{vanZee98}, \citet{Izotov99}, \citet{vanZee06} and \citet{Berg13}. In addition, we plot the Cl/O ratio from Milky Way \HII\ regions \citep{Esteban15} and from extra-galactic \HII\ regions from \citet{Izotov04} and \citet{Izotov06}.

All three elements, Ne, Ar and Cl appear to scale linearly with O. For Ne, the scaling agrees very well with that derived by \citet{Nieva12}  while for Cl, the scaling agrees very well with that determined by \citet{Esteban15}. A less good agreement is found for the SNRs in the case of Ar, with the SNR points falling systematically low by $\sim0.2-0.3$\,dex relative to the solar, \HII\ region and stellar determinations. The most likely explanation is that one or more of the Ar charge exchange reactions is in error, leading to a truncated \ion{Ar}{3} zone in these shock models.

\subsection{Elemental Abundances of the Magellanic Clouds}
\subsubsection{Large Magellanic Cloud}
Given that the SNRs add another method of determining interstellar abundances, it is useful to compare these with the recent stellar and \HII\ region determinations, given for  the LMC in Table \ref{table5}. Here we have drawn data for the B-stars from \citet{Hunter09,Korn02} and \citet{Korn05}. In addition we use data from F-supergiants from \citet{Andrievsky01}. For the \HII\ regions we use the work of \citet{Peimbert03, CarlosReyes15} and \citet{SanCipriano17} for abundances determined from collisionally-excited lines, while for recombination line abundances we use \citet{Peimbert03} and \citet{SanCipriano17}.

For the heavy elements found in refractory grains, we must rely on stellar abundance determinations. The mean of these  abundances were adopted in our models. Note that \citet{Peimbert03} gives the Fe abundance determined from the nebular [\ion{Fe}{3}] lines in 30 Doradus, which are approximately 1.0\,dex lower than the stellar Fe abundances, implying a logarithmic depletion of iron in the nebular gas of $\log D_{\mathrm{Fe}} = -1.0$. At this depletion, and using the depletion factors for other elements derived from the formulae of \citet{Jenkins09}, we infer $\log D_{\mathrm{C}} = -0.16$ and $\log D_{\mathrm{O}} = -0.02$. The figures given in Table \ref{table5} for \HII\ regions have been corrected for dust using the gas-phase abundances given in the various papers cited above.

The mean abundances for the LMC are given in the second column of Table 5. Given that the recombination line determinations are probably affected by resonant excitation of the permitted lines of C and O by the UV light of the exciting stars (as noted above), we have not used these in estimating the mean. In addition, given the offset noted above for the Ar abundance derived from SNRs, we have not included these in the averaging. The errors are estimated from the total scatter between the different determinations, rather from the internal errors given in the papers cited. 

Given the improvement in the observations and their method of analysis, the abundances given in Table \ref{table5} agree remarkably well with those derived by \citet{Russell92}. All elements agree within the stated errors, although the error bars in the current work are appreciably smaller than those given in the earlier work.

\begin{table*}
 \centering
 \small
   \caption{Estimated Mean Gas + Dust Elemental Abundances ($12+ \log X/H$) for the LMC}
    \label{table5}
   \scalebox{0.65}{
  \begin{tabular}{lcccccccccccc}
  \\
\hline
 &  & Hunter & Korn, & Korn & Andrievsky & This &  & Toribio San & Carlos Reyes & Peimbert & Toribio San & Peimbert\\
Element & $<{\mathrm{LMC}}>$ & 2009 & 2002 & 2005 & 2001 & Paper &  & Cipriano 2017 & 2015 & 2003 & Cipriano 2017 & 2003\\
&&&&&&&&&&&& \\
 &  & B-stars & B-stars & B-stars & F-stars & SNRs &  & HII regions & HII regions & 30 Dor & HII regions & 30 Dor\\
 &  &  &  &  &  &  &  & CELs & CELs & CELs & RLs & RLs\\
\hline
&&&&&&&&&&&& \\
He & $10.94\pm0.01$ & ... & 10.96 & ... & ... & 10.95 & ... & ... & 10.93 & ... & ... & 10.93\\
C & $8.02\pm0.10$ & 7.70 & 8.06 & 7.98 & 7.95 & 8.09 & ... & 8.08 & 8.17 & 8.08 & 8.12 & 8.31\\
N & $7.15\pm0.09$ & 7.13 & 7.01 & 7.02 & ... & 7.20 & ... & ... & 7.30 & 7.21 & ... & ...\\
O & $8.40\pm0.05$ & 8.34 & 8.37 & 8.40 & 8.51 & 8.34 & ... & 8.37 & 8.37 & 8.52 & 8.57 & 8.64\\
Ne & $7.70\pm0.09$ & ... & ... & ... & ... & 7.60 & ... & ... & 7.66 & 7.83 & ... & ...\\
S & $6.93\pm0.05$ & ... & ... & ... & ... & 6.98 & ... & ... & 6.83 & 6.99 & ... & ...\\
Cl & $4.89\pm0.07$ & ... & ... & ... & ... & 4.96 & ... & ... & ... & 4.82 & ... & ...\\
Ar & $6.17\pm0.09$ & ... & ... & ... & ... & 5.79 & ... & ... & 6.08 & 6.26 & ... & ...\\
&&&&&&&&&&&& \\
Mg & $7.19\pm0.09$ & 7.05 & 7.37 & 7.15 & ... & 7.19 & ... & ... & ... & ... & ... & ...\\
Ca & 6.02 & ... & ... & ... & ... & 6.02 & ... & ... & ... & ... & ... & ...\\
Si & $7.11\pm0.04$ & 7.17 & 7.10 & 7.07 & ... & 7.11 & ... & ... & ... & ... & ... & ...\\
Fe & 7.33 & ... & 7.33 & ... & ... & 7.33 & ... & ... & ... & (6.39) & ... & ...\\
Ni & 5.90 & ... & ... & ... & ... & 5.90 & ... & ... & ... & ... & ... & ...\\
&&&&&&&&&&&& \\
\hline
\multicolumn{13}{l}{{\citet{Peimbert03} value for Fe abundance is the measured gas-phase value.}} 
  \end{tabular}}
\end{table*}

\begin{table*}
 \centering
 \small
   \caption{Estimated Mean Gas + Dust Elemental Abundances ($12+ \log X/H$) for the SMC}
    \label{table6}
   \scalebox{0.58}{
  \begin{tabular}{lcccccccccccc}
  \\
\hline
Element:  &  &  Hunter & Bouret & This& Toribio San & Carlos Reyes & P\~ena-Guererro & Relano & Peimbert & Testor & Kurt & Toribio San\\
 &   $<{\mathrm{SMC}}>$ &  2009 & 2013 & Paper& Cipriano  2017 & 2015 & 2012 & 2002 & 2000 & 2001 & 1999 & Cipriano 2017\\
 &&&&&&&&&&&& \\
 &  &  B-stars & O-dwarfs & SNRs &  HII regions & HII regions & HII regions & HII regions & HII regions & HII regions & HII regions & HII regions\\
 &  &  &   &  &  CELs & CELs & CELs & CELs & CELs & CELs & CELs &  RLs\\
 \hline
 &&&&&&&&&&&& \\
He & $10.92\pm0.01$ & ... &  10.95 & 10.91 & ... &  10.92 & 10.92 & 10.90 & 10.89 & 10.92 & 10.91 & \\
C & $7.38\pm0.07$ & 7.30 & 7.30 & 7.50 & 7.23 & 7.42 & ... &  7.34 & ... & ... &  7.58 &  7.75\\
N &$ 6.81\pm0.23$ & 7.24 & 7.40 & 6.82 & ... &  6.56 & 6.72 & 6.81 & 6.51 & 6.64 & 6.59 & \\
O & $8.05\pm0.06$ & 7.99 & 7.97 & 8.02 & 8.01 & 8.00 & 8.08 & 8.16 & 8.16 & 8.00 & 8.06 & 8.33\\
Ne & $7.25\pm0.08$ & ... & ... &  7.04 & ... &  7.19 & 7.39 & 7.32 & 7.30 & 7.22 & 7.26 & \\
S & $6.48\pm0.13$ & ... & ... &  6.79 & ... &  6.32 & 6.53 & 6.40 & 6.59 & 6.31 & 6.42 & \\
Cl & $4.45\pm0.08$ & ... & ... &  4.37 & ... & ... &  4.52 & ... & ... &  ... & ... &  \\
Ar & $5.77\pm0.06$ & ... & ... &  5.71 & ... &  5.67 & 5.81 & 5.82 & 5.82 & 5.87 & 5.69 & \\
 &  &  &   &  &   &  &   &  &   &  &  &  \\
Mg & 6.72 & 6.72 & ... &  6.72 & ... & ... &  ... & ... &  ... & ... &  ... &  \\
Ca & 5.77 & ... & ... &  5.77 & ... & ... &  ... & ... &  ... & ... &  ... &  \\
Si & 6.75 & 6.77 & ... &  6.77 & ... & ... &  ... & ... &  ... & 6.70 &  ... &  \\
Fe & 6.77 & ... & ... &  6.77 & ... & ... &  ... &  (6.16) & ... & ... &  ... &  \\
Ni & 5.55 & ... & ... &  5.55 & ... & ... &  ... & ... &  ... & ... &  ... &  \\
&&&&&&&&&&&& \\
\hline
 \multicolumn{13}{l}{{\citet{Relano02} Fe abundance in parenthesis is the measured gas-phase value.}} 
  \end{tabular}}
\end{table*}

\subsubsection{Small Magellanic Cloud}
For the SMC we use the data from  \citet{Hunter09} for B Type stars, and \citet{Bouret13} for O Type Dwarf stars. For the \HII\ regions, we use the work by \citet{SanCipriano17, CarlosReyes15, Pena-Guerrero12, Relano02, Peimbert00, Testor01} and \citet{Kurt99}. The data from the  [\ion{Fe}{3}] lines reported by \citet{Relano02} compared with the Fe abundance derived by \citet{Hunter09} implies  $\log D_{\mathrm{Fe}} = -0.6$. At this depletion, we expect $\log D_{\mathrm{C}} = -0.06$ and $\log D_{\mathrm{O}} = -0.01$.  The \HII\ region values given in Table \ref{table6} have been corrected by this amount. Once again, the recombination line abundances for the \HII\ regions are systematically high, and have not been used in the averaging.

\section{Conclusions}
Although the idea of measuring abundances and abundance gradients in the ISM using radiative supernova shockwaves is not new \citep{Dopita77, Dopita80}, the advances in instrumental capability, analysis and shockwave modelling warrant a re-examination of this technique. In this paper we have provided integral field spectroscopy and analysis based upon L1-norm minimisation which establishes that the technique can provide chemical abundances with a precision comparable to those given by the analysis of B-type stars, or from the analysis of \HII\ regions.

It is interesting that the fast shocks which generate the majority of the line emission in the SNRs studied here all have shock velocities in the range $200-230$\,km/s. { We have shown that the most luminous shock will be the fastest fully radiative shock and at a shock velocity of  $200-230$\,km/s, the plasma is heated to the peak of the cooling function, so in general this velocity will correspond to the fastest radiative shock. However, it should also be noted that } such shocks are marginally thermally-unstable, so the plane-parallel steady-flow models used here become somewhat questionable. The effect of thermal instabilities in a 3-D geometry is to break up the flow, and to generate weak secondary shocks \citep{Sutherland03} which may collectively significantly contribute to the overall emission. This may be the physical reason that our model fits were much improved by the addition of a slow-shock component.

We have shown that the radiative knots in Kepler are dense interstellar clouds which have been recently shocked, rather than representing material ejected by the pre-supernova star. The abundance gradient inferred for our Galaxy is compatible with that found in B-type stars. For the LMC and the SMC we have combined recent data taken from all available techniques to provide new estimates of the chemical abundances in these systems. 

It seems clear that the abundances delivered by the recombination lines of C and O are systematically higher than those give by other techniques, even when the depletion of these elements onto dust grains is taken into account. The most likely explanation is that the intensity of these lines is enhanced by UV fluorescent processes driven by the UV radiation field of the central stars.

For the refractory elements normally trapped in dust grains, our results are generally consistent with the conclusions of \citep{Dopita16} and \citep{Dopita18} that dust has been mostly destroyed in the region emitting the [\ion{Fe}{2}] lines, while a smaller fraction has been destroyed in the \ion{Fe}{3} and \ion{Fe}{7} zones, consistent with the models of \citet{Seab83} and \citet{Borkowski95}. However, in Kepler, virtually all dust has been destroyed. Dust destruction is also rather advanced in IC 1613 S8.

\section*{Acknowledgments}
MD and RS acknowledge the support of the Australian Research Council (ARC) through Discovery project DP16010363. Parts of this research were conducted by the Australian Research Council Centre of Excellence for All Sky Astrophysics in 3 Dimensions (ASTRO 3D), through project number CE170100013. FPAV and IRS thank the CAASTRO AI travel grant for generous support. IRS was supported by the ARC through the Future Fellowship grant FT1601000028. AJR is supported by the Australian Research Council through Future Fellowship grant FT170100243.


\bibliographystyle{aasjournal}

\newpage
\setcounter{table}{0}
\renewcommand{\thetable}{A\arabic{table}}
 \section*{Appendix}

\begin{table*}
 \centering
 \small
   \caption{Measured line fluxes for Kepler's SNR}
    \label{tableA1}
   \scalebox{0.45}{
  \begin{tabular}{lccccc}
  \\
 \hline
 &  & Kepler 1 &  & Kepler 2 & \\
 $\lambda$ (\AA)  & ID &  De-red flux  & Error &  De-red flux  & Error \\
 \hline
3726.0 & [O II] & 311.1 & 17.5 & 270.6 & 12.6 \\
3728.8 & [O II] & 110.8 & 16.6 & 91.7 & 10.5 \\
3835.4 & HI & -- & -- & -- & -- \\
3869.8 & [Ne III] & 75.2 & 5.4 & 75.6 & 5.0 \\
3888.6 & HeI, HI & 24.8 & 5.4 & 14.4 & 5.3 \\
3933.7 & Ca II & 25.5 & 4.0 & 27.4 & 3.9 \\
3969.0 & [Ne III], Ca II, H$\epsilon$ & 58.5 & 11.5 & 46.9 & 10.8 \\
4026.0 & He I & -- & -- & -- & -- \\
4068.6 & [S II] & 124.2 & 2.9 & 134.2 & 3.5 \\
4076.3 & [S II] & 33.2 & 2.6 & 56.8 & 3.1 \\
4101.7 & H$\delta$ & 20.0 & 2.8 & 32.2 & 2.9 \\
4244.8 & [Fe II] & 7.2 & 1.5 & 6.7 & 3.6 \\
4276.8 & [Fe II] & 2.2 & 1.3 & -- & -- \\
4287.4 & [Fe II] & 3.4 & 0.9 & 7.7 & 1.4 \\
4319.6 & [Fe II] & 4.1 & 1.4 & 1.5 & 1.1 \\
4340.5 & H$\gamma$ & 47.7 & 1.6 & 47.5 & 1.5 \\
4359.1 & [Fe IX] & 11.7 & 2.3 & 9.8 & 1.4 \\
4363.2 & [O III] & 12.9 & 1.8 & 10.6 & 1.6 \\
4416.3 & [Fe II] & 11.1 & 2.1 & 14.7 & 2.5 \\
4452.1 & [Fe II] & 1.6 & 0.7 & 3.0 & 1.1 \\
4458.0 & [Fe III] & 2.1 & 0.8 & 5.1 & 1.5 \\
4471.5 & He I & 5.5 & 2.2 & 5.0 & 1.2 \\
4566.8 & Mg I] & 8.1 & 1.1 & 10.2 & 0.9 \\
4658.1 & [Fe III] & 15.3 & 1.1 & 12.8 & 0.9 \\
4685.7 & He II & 3.8 & 1.0 & 5.5 & 0.8 \\
4701.5 & [Fe II] & 6.3 & 1.1 & 5.2 & 0.8 \\
4754.7 & [Fe III] & 3.5 & 1.1 & 2.1 & 0.5 \\
4814.5 & [Fe III] & 3.1 & 0.7 & 6.6 & 1.3 \\
4861.3 & H$\beta$ & 100.0 & 1.2 & 100.0 & 1.1 \\
4881.0 & [Fe III] & 8.5 & 1.3 & 6.2 & 0.9 \\
4905.4 & [Fe II] & 2.9 & 0.6 & 2.6 & 1.4 \\
4921.9 & He I & 0.7 & 0.8 & 1.8 & 0.6 \\
4958.9 & [O III] & 61.2 & 1.0 & 50.8 & 1.0 \\
5006.8 & [O III] & 191.1 & 1.4 & 155.6 & 1.5 \\
5015.7 & He I & 3.5 & 0.9 & 4.9 & 2.1 \\
5111.6 & [Fe II] & 2.2 & 0.4 & 3.7 & 0.7 \\
5158.8 & [Fe II], [Fe VII] & 24.7 & 0.7 & 25.2 & 0.6 \\
5199.5 & [N I] & 11.6 & 0.7 & 8.5 & 0.6 \\
5220.1 & [Fe II] & 0.6 & 0.8 & 2.5 & 0.5 \\
5261.6 & [Fe II] & 9.8 & 0.7 & 10.8 & 0.6 \\
5271.0 & [Fe III], [Fe II] & 4.9 & 1.3 & 10.0 & 1.0 \\
5303.3 & [Fe XIV] & 6.0 & 1.3 & 3.3 & 1.2 \\
5333.7 & [Fe II], [Fe VI] & 6.1 & 0.6 & 7.2 & 0.6 \\
5376.4 & [Fe II] & 2.5 & 0.6 & 4.0 & 0.6 \\
5527.3 & [Fe II] & 5.3 & 1.0 & 7.3 & 0.6 \\
5754.6 & [N II] & 21.5 & 0.6 & 21.7 & 0.4 \\
5875.6 & He I & 18.8 & 0.6 & 17.4 & 0.4 \\
6087.0 & [Fe VII] & 2.6 & 0.6 & 1.2 & 0.3 \\
6300.3 & [O I] & 144.2 & 0.9 & 149.7 & 0.8 \\
6312.1 & [S III] & 2.6 & 0.6 & 2.5 & 0.5 \\
6363.8 & [O I] & 48.6 & 0.4 & 48.5 & 0.3 \\
6374.5 & [Fe X] & 1.5 & 0.3 & 1.9 & 0.4 \\
6548.0 & [N II] & 227.2 & 2.7 & 187.4 & 1.7 \\
6562.8 & H$\alpha$ & 308.1 & 2.8 & 301.0 & 2.0 \\
6584.5 & [N II] & 692.8 & 4.1 & 494.7 & 2.8 \\
6678.1 & He I & 3.6 & 0.3 & 3.5 & 0.5 \\
6716.4 & [S II] & 85.5 & 1.1 & 68.8 & 0.7 \\
6730.8 & [S II] & 174.5 & 1.1 & 140.6 & 0.9 \\
7065.2 & He I & 6.1 & 0.2 & 5.8 & 0.1 \\
7135.8 & [Ar III] & 12.1 & 0.3 & 10.0 & 0.2 \\
7154.8 & [Fe II] & 39.5 & 0.3 & 38.7 & 0.2 \\
7171.6 & [Fe II] & 10.2 & 0.2 & 10.4 & 0.1 \\
7291.5 & [Ca II] & 57.0 & 0.6 & 60.4 & 0.5 \\
7321.0 & [Ca II], [O II] & 103.1 & 1.2 & 100.6 & 1.7 \\
7330.1 & [O II] & 49.7 & 1.0 & 43.0 & 1.2 \\
7377.8 & [Ni II] & 34.2 & 0.4 & 33.4 & 0.4 \\
7388.2 & [Fe II] & 7.3 & 0.4 & 7.5 & 0.3 \\
7411.6 & [Ni II] & 3.7 & 0.4 & 3.5 & 0.3 \\
7452.1 & [Fe II] & 13.3 & 0.4 & 12.0 & 0.3 \\
7631.5 & [Fe II] & 6.6 & 0.3 & 6.6 & 0.3 \\
7665.3 & [Fe II] & 2.2 & 0.4 & 2.6 & 0.3 \\
7686.9 & [Fe II] & 3.5 & 0.3 & 4.2 & 0.2 \\
7891.9 & [Fe XI] & 1.0 & 0.2 & 2.3 & 0.3 \\
8125.5 & [Cr II] & 2.5 & 0.2 & 3.1 & 0.2 \\
8229.8 & [Cr II] & 2.3 & 0.2 & 2.3 & 0.1 \\
8234.5 & [Fe IX] & 1.1 & 0.2 & 1.5 & 0.1 \\
8542.1 & [Ca II] & 3.8 & 0.4 & 2.7 & 0.2 \\
8578.7 & [Cl II] & 5.2 & 0.3 & 3.8 & 0.2 \\
8616.9 & [Fe II] & 46.2 & 0.5 & 41.8 & 0.3 \\
8662.1 & [Ca II] & 1.8 & 0.4 & 1.5 & 0.4 \\
8727.1 & [C I] & 3.2 & 0.1 & 3.2 & 0.1 \\
8891.9 & [Fe II] & 13.1 & 0.7 & 14.5 & 0.5 \\
\hline
FWHM(km/s)  & & 186 &  & 151 & \\
 c & & 1.30 &  & 1.70 & \\
 $S_{H\beta}$ & & 8.01E-15 &  & 2.22E-14 & \\
 \hline
  \end{tabular}}
\end{table*}

\begin{table*}
 \centering
 \small
   \caption{Measured line fluxes for SNRs in the LMC}
    \label{tableA2}
   \scalebox{0.45}{
  \begin{tabular}{lccccccc}
  \\
 \hline
 & & N132D P14 &  & N49 S  & & N103B C & \\
 $\lambda$ (\AA) & ID &  De-red flux  & Error. &  De-red flux  & Error. &  De-red flux  & Error.\\
 \hline
3726.0 & [O II] & 229.5 & 2.0 & 360.2 & 10.8 & 101.0 & 10.9\\
3728.8 & [O II] & 119.5 & 2.0 & 202.2 & 10.0 & 71.0 & 11.2\\
3835.4 & HI & 5.47 & 0.45 & 9.0 & 1.2 & 6.5 & 0.5\\
3869.8 & [Ne III] & 32.2 & 0.5 & 37.0 & 1.2 & 31.0 & 0.7\\
3888.6 & HeI, HI & 19.8 & 0.4 & 22.1 & 0.9 & 22.8 & 0.5\\
3933.7 & Ca II & 16.8 & 0.4 & 15.6 & 1.3 & 19.3 & 0.5\\
3969.0 & [Ne III], Ca II, H$\epsilon$ & 32.5 & 0.6 & 33.3 & 1.7 & 33.2 & 0.7\\
4026.0 & He I & 2.4 & 0.4 & 2.1 & 0.6 & 2.4 & 0.4\\
4068.6 & [S II] & 41.4 & 0.3 & 29.6 & 0.8 & 46.2 & 0.9\\
4076.3 & [S II] & 14.1 & 0.3 & 9.7 & 0.8 & 15.6 & 0.9\\
4101.7 & H$\delta$ & 24.5 & 0.3 & 25.2 & 0.6 & 25.3 & 0.5\\
4244.8 & [Fe II] & 4.9 & 0.2 & 4.7 & 0.4 & 5.2 & 0.2\\
4276.8 & [Fe II] & 1.5 & 0.2 & 1.3 & 0.4 & 1.7 & 0.3\\
4287.4 & [Fe II] & 3.9 & 0.2 & 4.2 & 0.4 & 4.8 & 0.3\\
4319.6 & [Fe II] & 0.7 & 0.4 & 0.5 & 0.3 & 0.7 & 0.2\\
4340.5 & H$\gamma$ & 44.8 & 0.4 & 45.4 & 1.2 & 46.3 & 0.8\\
4359.1 & [Fe IX] & 4.1 & 0.4 & 5.0 & 2.6 & 5.6 & 1.0\\
4363.2 & [O III] & 7.1 & 0.3 & 3.9 & 1.6 & 5.9 & 0.8\\
4416.3 & [Fe II] & 4.9 & 0.3 & 4.7 & 0.8 & 6.2 & 0.5\\
4452.1 & [Fe II] & 1.0 & 0.1 & 1.2 & 0.4 & 1.5 & 0.2\\
4458.0 & [Fe III] & 0.9 & 0.2 & 1.0 & 0.3 & 1.4 & 0.2\\
4471.5 & He I & 3.9 & 0.2 & 4.1 & 0.4 & 4.8 & 0.3\\
4566.8 & Mg I] & 3.6 & 0.2 & 2.6 & 0.4 & 3.8 & 0.3\\
4658.1 & [Fe III] & 5.1 & 0.2 & 4.6 & 0.3 & 3.9 & 0.2\\
4685.7 & He II & 5.5 & 0.2 & 5.4 & 0.4 & 5.0 & 0.3\\
4701.5 & [Fe II] & 1.1 & 0.1 & 1.7 & 1.1 & 1.4 & 0.5\\
4754.7 & [Fe III] & 0.6 & 0.2 & 0.8 & 0.3 & 0.5 & 0.2\\
4814.5 & [Fe III] & 2.4 & 0.2 & 2.9 & 0.3 & 2.7 & 0.1\\
4861.3 & H$\beta$ & 100.0 & 0.3 & 100.0 & 2.5 & 100.0 & 1.6\\
4881.0 & [Fe III] & 2.1 & 0.2 & 1.8 & 0.7 & 1.6 & 0.7\\
4905.4 & [Fe II] & 0.9 & 0.2 & 0.8 & 0.6 & 1.1 & 0.6\\
4921.9 & He I & -- & -- & 1.2 & 0.6 & 1.0 & 0.5\\
4958.9 & [O III] & 38.4 & 0.3 & 25.5 & 1.0 & 28.0 & 1.2\\
5006.8 & [O III] & 114.9 & 0.8 & 78.3 & 2.0 & 84.4 & 2.6\\
5015.7 & He I & -- & -- & 1.2 & 0.2 & 1.6 & 0.2\\
5111.6 & [Fe II] & 1.6 & 0.1 & 11.9 & 0.5 & 12.0 & 0.4\\
5158.8 & [Fe II], [Fe VII] & 11.7 & 0.2 & 3.5 & 0.4 & 1.7 & 0.2\\
5199.5 & [N I] & 3.1 & 0.3 & 0.7 & 0.2 & 1.0 & 0.1\\
5220.1 & [Fe II] & -- & -- & 3.9 & 0.3 & 4.9 & 0.2\\
5261.6 & [Fe II] & 4.1 & 0.1 & 4.5 & 1.0 & 4.5 & 0.6\\
5271.0 & [Fe III], [Fe II] & 0.9 & 1.0 & 3.2 & 1.2 & 0.0 & 0.4\\
5303.3 & [Fe XIV] & 3.7 & 0.2 & 1.2 & 0.3 & 2.9 & 0.4\\
5333.7 & [Fe II], [Fe VI] & 2.2 & 0.1 & 1.8 & 0.3 & 2.7 & 0.2\\
5376.4 & [Fe II] & 1.1 & 0.2 & 0.8 & 0.3 & 1.7 & 0.2\\
5527.3 & [Fe II] & 2.6 & 0.3 & 2.3 & 0.3 & 2.8 & 0.4\\
5754.6 & [N II] & 1.3 & 0.1 & 1.3 & 0.3 & 2.0 & 0.2\\
5875.6 & He I & 11.5 & 0.1 & 11.2 & 0.5 & 12.5 & 0.2\\
6087.0 & [Fe VII] & 0.6 & 0.1 & 0.4 & 0.2 & 0.6 & 0.1\\
6300.3 & [O I] & 104.5 & 1.3 & 90.0 & 8.3 & 94.2 & 2.3\\
6312.1 & [S III] & 0.9 & 0.4 & 1.0 & 7.0 & 1.0 & 4.3\\
6363.8 & [O I] & 35.2 & 0.3 & 30.9 & 3.4 & 34.2 & 0.7\\
6374.5 & [Fe X] & 2.7 & 0.3 & 1.1 & 2.3 & 2.2 & 1.0\\
6548.0 & [N II] & 29.5 & 1.0 & 21.9 & 3.5 & 18.0 & 7.3\\
6562.8 & H$\alpha$ & 305.4 & 1.3 & 301.1 & 16.5 & 294.0 & 8.0\\
6584.5 & [N II] & 91.4 & 0.4 & 59.2 & 5.3 & 54.8 & 1.9\\
6678.1 & He I & 3.2 & 0.1 & 2.8 & 0.2 & 2.9 & 0.2\\
6716.4 & [S II] & 85.8 & 0.3 & 85.1 & 10.4 & 21.8 & 0.7\\
6730.8 & [S II] & 144.9 & 0.4 & 120.6 & 11.3 & 44.9 & 0.7\\
7065.2 & He I & 5.2 & 0.1 & -- & -- & -- & --\\
7135.8 & [Ar III] & 4.6 & 0.1 & -- & -- & -- & --\\
7154.8 & [Fe II] & 22.1 & 0.1 & -- & -- & -- & --\\
7171.6 & [Fe II] & 4.7 & 0.1 & -- & -- & -- & --\\
7291.5 & [Ca II] & 39.1 & 0.3 & -- & -- & -- & --\\
7321.0 & [Ca II], [O II] & 51.9 & 0.8 & -- & -- & -- & --\\
7330.1 & [O II] & 18.8 & 0.5 & -- & -- & -- & --\\
7377.8 & [Ni II] & 10.2 & 0.1 & -- & -- & -- & --\\
7388.2 & [Fe II] & 3.2 & 0.1 & -- & -- & -- & --\\
7411.6 & [Ni II] & 0.6 & 0.1 & -- & -- & -- & --\\
7452.1 & [Fe II] & 7.0 & 0.1 & -- & -- & -- & --\\
7631.5 & [Fe II] & 4.3 & 0.1 & -- & -- & -- & --\\
7665.3 & [Fe II] & 0.8 & 0.1 & -- & -- & -- & --\\
7686.9 & [Fe II] & 2.0 & 0.1 & -- & -- & -- & --\\
7891.9 & [Fe XI] & 2.8 & 0.3 & -- & -- & -- & --\\
8125.5 & [Cr II] & 1.6 & 0.1 & -- & -- & -- & --\\
8229.8 & [Cr II] & 1.3 & 0.1 & -- & -- & -- & --\\
8234.5 & [Fe IX] & 2.0 & 0.1 & -- & -- & -- & --\\
8542.1 & [Ca II] & 2.1 & 0.1 & -- & -- & -- & --\\
8578.7 & [Cl II] & 1.8 & 0.1 & -- & -- & -- & --\\
8616.9 & [Fe II] & 27.9 & 0.3 & -- & -- & -- & --\\
8662.1 & [Ca II] & 2.5 & 0.4 & -- & -- & -- & --\\
8727.1 & [C I] & 1.2 & 0.1 & -- & -- & -- & --\\
8891.9 & [Fe II] & 7.5 & 0.4 & -- & -- & -- & --\\
\hline 
FWHM(km/s) & & 183 & & 165 & & 224 & \\
c & & 0.00 & & 0.70 & & 0.48 & \\
$S_{H\beta}$ & & 1.99E-15 &  & 1.32E-14 & & 1.43E-14 &  \\
 \hline
  \end{tabular}}
\end{table*}

\begin{table*}
 \centering
 \small
   \caption{Measured line fluxes for SNRs in other galaxies}
    \label{tableA3}
   \scalebox{0.45}{
  \begin{tabular}{lccccccc}
  \\
 \hline
 &  & SMC: &  & NGC6822: &  & IC1613: & \\
 &  & SNR0104-72.3 &  & Ho 12 &  & S 8 & \\
 $\lambda$ (\AA) & ID &  De-red flux  & Error. &  De-red flux  & Error. &  De-red flux  & Error.\\
\hline
3726.0 & [O II] & 74.5 & 2.5 & 131.6 & 13.4 & 89.9 & 6.0\\
3728.8 & [O II] & 78.4 & 2.4 & 129.6 & 11.8 & 58.3 & 6.1\\
3835.4 & HI & 4.5 & 1.0 & -- & -- & 5.2 & 0.7\\
3869.8 & [Ne III] & 4.3 & 0.9 & 16.3 & 3.4 & 9.1 & 1.2\\
3888.6 & HeI, HI & 12.1 & 1.1 & 19.2 & 4.8 & 19.0 & 1.1\\
3933.7 & Ca II & 3.7 & 0.6 & 14.8 & 4.0 & 10.6 & 1.0\\
3969.0 & [Ne III], Ca II, H$\epsilon$ & 14.9 & 1.0 & 24.4 & 2.8 & 22.6 & 1.4\\
4026.0 & He I & -- & -- & -- & -- & 1.8 & 0.7\\
4068.6 & [S II] & 6.5 & 0.6 & 11.0 & 1.8 & 12.6 & 0.6\\
4076.3 & [S II] & 2.9 & 0.8 & 4.4 & 0.6 & 5.1 & 1.1\\
4101.7 & H$\delta$ & 21.0 & 0.5 & 28.7 & 2.1 & 24.0 & 0.7\\
4244.8 & [Fe II] & -- & -- & -- & -- & 2.6 & 0.5\\
4276.8 & [Fe II] & -- & -- & -- & -- & 0.6 & 0.3\\
4287.4 & [Fe II] & -- & -- & 5.5 & 1.0 & 2.1 & 0.4\\
4319.6 & [Fe II] & -- & -- & -- & -- & -- & --\\
4340.5 & H$\gamma$ & 42.6 & 0.5 & 43.3 & 1.0 & 46.2 & 0.5\\
4359.1 & [Fe IX] & -- & -- & -- & -- & 1.8 & 0.3\\
4363.2 & [O III] & 1.2 & 0.6 & 4.9 & 0.9 & 3.1 & 0.5\\
4416.3 & [Fe II] & 1.2 & 0.5 & 7.0 & 3.9 & 3.3 & 0.5\\
4452.1 & [Fe II] & -- & -- & -- & -- & 1.1 & 0.5\\
4458.0 & [Fe III] & 0.4 & 0.1 & -- & -- & -- & --\\
4471.5 & He I & 2.4 & 0.3 & -- & -- & 3.1 & 0.4\\
4566.8 & Mg I] & 1.8 & 0.4 & 2.0 & 0.7 & 1.5 & 0.3\\
4658.1 & [Fe III] & 1.4 & 0.4 & 1.0 & 0.6 & 1.7 & 0.3\\
4685.7 & He II & 3.4 & 0.4 & 4.0 & 0.9 & 5.9 & 0.3\\
4701.5 & [Fe II] & 0.1 & 0.0 & -- & -- & 3.2 & 2.7\\
4754.7 & [Fe III] & 0.5 & 0.2 & -- & -- & 0.5 & 0.2\\
4814.5 & [Fe III] & 0.9 & 0.7 & -- & -- & 1.7 & 0.3\\
4861.3 & H$\beta$ & 100.0 & 0.5 & 100.0 & 1.1 & 100.0 & 0.8\\
4881.0 & [Fe III] & -- & -- & 1.6 & 0.8 & 1.0 & 0.4\\
4905.4 & [Fe II] & -- & -- & -- & -- & 0.6 & 0.3\\
4921.9 & He I & 0.9 & 0.2 & 2.3 & 0.6 & 1.0 & 0.5\\
4958.9 & [O III] & 7.1 & 0.4 & 23.2 & 0.7 & 11.8 & 0.4\\
5006.8 & [O III] & 20.4 & 0.7 & 69.0 & 0.9 & 35.3 & 1.0\\
5015.7 & He I & 1.0 & 0.3 & -- & -- & 0.9 & 0.4\\
5111.6 & [Fe II] & -- & -- & -- & -- & 0.9 & 0.3\\
5158.8 & [Fe II], [Fe VII] & 2.3 & 0.3 & 8.1 & 0.8 & 6.8 & 0.4\\
5199.5 & [N I] & 2.7 & 0.5 & 1.6 & 0.7 & 1.4 & 0.1\\
5220.1 & [Fe II] & -- & -- & -- & -- & 0.6 & 0.3\\
5261.6 & [Fe II] & 0.7 & 0.3 & 3.4 & 0.6 & 2.5 & 0.3\\
5271.0 & [Fe III], [Fe II] & -- & -- & 5.3 & 1.3 & 0.4 & 0.5\\
5303.3 & [Fe XIV] & 5.8 & 3.5 & 3.1 & 0.7 & 1.9 & 0.5\\
5333.7 & [Fe II], [Fe VI] & 0.2 & 0.1 & 0.5 & 0.3 & 0.9 & 0.5\\
5376.4 & [Fe II] & -- & -- & -- & -- & 0.3 & 0.2\\
5527.3 & [Fe II] & -- & -- & -- & -- & 1.1 & 0.6\\
5754.6 & [N II] & 1.9 & 0.6 & 2.6 & 0.9 & 1.1 & 0.9\\
5875.6 & He I & 8.7 & 0.3 & 9.4 & 0.3 & 10.9 & 0.3\\
6087.0 & [Fe VII] & 0.5 & 0.1 & 1.5 & 0.2 & 1.4 & 0.4\\
6300.3 & [O I] & 49.1 & 0.7 & 58.0 & 1.4 & 38.2 & 0.4\\
6312.1 & [S III] & 0.3 & 0.1 & -- & -- & 1.0 & 0.7\\
6363.8 & [O I] & 16.5 & 0.3 & 18.7 & 0.7 & 13.3 & 0.4\\
6374.5 & [Fe X] & 0.4 & 0.2 & -- & -- & 0.5 & 0.4\\
6548.0 & [N II] & 4.4 & 0.3 & 11.3 & 1.4 & 7.1 & 1.0\\
6562.8 & H$\alpha$ & 342.8 & 1.2 & 338.7 & 1.9 & 331.0 & 2.3\\
6584.5 & [N II] & 16.3 & 0.4 & 33.9 & 1.0 & 20.4 & 0.6\\
6678.1 & He I & 3.2 & 0.3 & 2.7 & 0.1 & 2.8 & 0.2\\
6716.4 & [S II] & 83.4 & 0.5 & 83.8 & 1.1 & 54.5 & 0.3\\
6730.8 & [S II] & 60.8 & 0.4 & 97.1 & 1.2 & 74.1 & 0.3\\
7065.2 & He I & 2.6 & 0.3 & 3.5 & 0.2 & 3.7 & 0.2\\
7135.8 & [Ar III] & -- & -- & 2.3 & 0.2 & 1.4 & 0.2\\
7154.8 & [Fe II] & -- & -- & 8.9 & 0.2 & 8.5 & 0.2\\
7171.6 & [Fe II] & -- & -- & 1.2 & 0.1 & 1.0 & 0.1\\
7291.5 & [Ca II] & 12.5 & 0.3 & 21.7 & 0.6 & 19.0 & 0.5\\
7321.0 & [Ca II], [O II] & 9.9 & 0.7 & 25.2 & 0.5 & 18.6 & 1.5\\
7330.1 & [O II] & 1.5 & 0.3 & 5.8 & 0.3 & 5.0 & 1.4\\
7377.8 & [Ni II] & 0.8 & 0.2 & 4.0 & 0.2 & 3.4 & 0.3\\
7388.2 & [Fe II] & -- & -- & -- & -- & 1.2 & 0.3\\
7411.6 & [Ni II] & -- & -- & -- & -- & 0.2 & 0.1\\
7452.1 & [Fe II] & -- & -- & 3.3 & 0.2 & 2.8 & 0.3\\
7631.5 & [Fe II] & 1.5 & 0.4 & 1.2 & 0.2 & 1.1 & 0.3\\
7665.3 & [Fe II] & 0.2 & 0.2 & -- & -- & 0.5 & 0.4\\
7686.9 & [Fe II] & 0.4 & 0.1 & 0.6 & 0.2 & 0.8 & 0.2\\
7891.9 & [Fe XI] & -- & -- & 2.3 & 0.4 & 1.0 & 0.3\\
8125.5 & [Cr II] & -- & -- & 0.6 & 0.2 & 0.7 & 0.2\\
8229.8 & [Cr II] & -- & -- & -- & -- & -- & --\\
8234.5 & [Fe IX] & -- & -- & 2.2 & 0.4 & 1.4 & 0.3\\
8542.1 & [Ca II] & -- & -- & 1.9 & 0.4 & 1.1 & 0.4\\
8578.7 & [Cl II] & 0.4 & 0.1 & -- & -- & -- & --\\
8616.9 & [Fe II] & 3.6 & 0.5 & 12.5 & 0.5 & 9.7 & 0.3\\
8662.1 & [Ca II] & 0.9 & 0.3 & 1.7 & 0.2 & 2.1 & 0.9\\
8727.1 & [C I] & 0.8 & 0.2 & -- & -- & -- & --\\
8891.9 & [Fe II] & 0.8 & 0.3 & 3.8 & 0.6 & 3.1 & 1.2\\
\hline
FWHM(km/s) &  & 126 &  & 225 &  & 231 & \\
c &  & 0.25 &  & 0.39 &  & 0.10 & \\
$S_{H\beta}$ &  & 1.50E-15 &  & 6.21E-16 &  & 6.91E-15 & \\
\hline
  \end{tabular}}
\end{table*}

\end{document}